\documentclass{lmcs} 
\usepackage[utf8]{inputenc}
\pdfoutput=1

\usepackage{lastpage}
\lmcsdoi{19}{1}{18}
\lmcsheading{}{\pageref{LastPage}}{}{}%
{Dec.~20,~2021}{Mar.~07,~2023}{}


\usepackage{hyperref}
\usepackage{mathtools}
\usepackage{thmtools}
\usepackage{thm-restate}
\usepackage{nicefrac}
\usepackage{subcaption}

 \newcommand{\review}[1]{#1} 

\newcommand{\exaend}{\hfill$\lhd$}

\DeclarePairedDelimiter{\ceil}{\lceil}{\rceil}
\allowdisplaybreaks%
\usetikzlibrary{automata,arrows,calc}
\tikzset{->,>=stealth',auto,node distance=2cm,thick,initial text=}
\tikzstyle{accepting}=[path picture={%
  \draw let
    \p1 = (path picture bounding box.east),
    \p2 = (path picture bounding box.center)
    in
    (\p2) circle (\x1 - \x2 - 2pt); 
 }]
\definecolor{green1}{rgb}{0, 0.5, 0}
\definecolor{red1}{rgb}{0.64, 0, 0}

\newcommand{\R}{\mathbb{R}} 
\newcommand{\IfunM}{\Inter_{min}} 
\newcommand{\mx}[1]{max^{#1}} 
\newcommand{\mn}[1]{min^{#1}} 
\newcommand{\Path}[2]{\infl{#1}{#2}} 
\newcommand{\CBfactor}{\beta}
\newcommand{\CBfun}[3]{\CBfactor^{#3}_{\agent{#1},\agent{#2}}} 
\newcommand{\CBfunM}{\CBfactor_{min}}

\newcommand{\Agents}{\mathcal{A}} 
\newcommand{\agent}[1]{#1} 
\newcommand{\Blf}{B} 
\newcommand{\Blft}[1]{\Blf^{#1}} 
\newcommand{\Bfun}[2]{\Blf^{#2}_{\agent{#1}}} 

\newcommand{\qm}[1]{``#1''}

\newcommand{\cali}{\mathcal{I}}

\newcommand{\calt}{\mathcal{T}}

\newcommand{\Pol}{\rho}
\newcommand{\Pfun}[1]{\Pol(#1)}
\newcommand{\PolER}{\rho_{\mathit{ER}}}
\newcommand{\PfunER}[1]{\PolER(#1)}
\newcommand{\BD}[2]{\ensuremath{\mathbf{bd}(#1,#2)}}

\newcommand{\Inter}{\cali} 
\newcommand{\Interclique}{\Inter^{\textit{clique}}} 
\newcommand{\Interdisconnected}{\Inter^{\textit{disc}}} 
\newcommand{\Interfaint}{\Inter^{\textit{faint}}} 
\newcommand{\Interunrelenting}{\Inter^{\textit{unrel}}} 
\newcommand{\Intermalleable}{\Inter^{\textit{malleable}}} 
\newcommand{\Intercircular}{\Inter^{\textit{circ}}} 

\newcommand{\Ifun}[2]{\Inter_{#1,#2}} 
\newcommand{\Ifunclique}[2]{\Interclique_{#1,#2}} 
\newcommand{\Ifundisconnected}[2]{\Interdisconnected_{#1,#2}} 
\newcommand{\Ifunfaint}[2]{\Interfaint_{#1,#2}} 
\newcommand{\Ifununrelenting}[2]{\Interunrelenting_{#1,#2}} 
\newcommand{\Ifunmalleable}[2]{\Intermalleable_{#1,#2}} 
\newcommand{\Ifuncircular}[2]{\Intercircular_{#1,#2}} 

\newcommand{\Upd}{\mu} 

\newcommand{\UpdR}{\mu^{C}} 

\newcommand{\UpdCB}{\mu^{\textit{CB}}} 

\newcommand{\larrow}[1]{\stackrel{\,\, #1\,\,}{\rightarrow}} 
\newcommand{\Larrow}[2]{\stackrel{\,\,#1\,\,}{\leadsto_{#2}}}
\newcommand{\ldinfl}[3]{#1{\larrow{#2}}#3} 
\newcommand{\infl}[2]{#1{\Larrow{}{}}#2} 
\newcommand{\linfl}[4]{#1 \Larrow{#2}{#3} #4} 

\newcommand{\nat}{\mathbb{N}}
\newcommand{\reals}{\mathbb{R}}
\newcommand{\mstar}{\mathtt{m}}

\newcommand{\defsymbol}{\stackrel{\textup{\texttt{def}}}  {=}}

\makeatletter

\newdimen\w@dth

\def\setw@dth#1#2{\setbox\z@\hbox{\scriptsize $#1$}\w@dth=\wd\z@
\setbox\@ne\hbox{\scriptsize $#2$}\ifnum\w@dth<\wd\@ne \w@dth=\wd\@ne \fi
\advance\w@dth by 1.2em}

\def\t@^#1_#2{\allowbreak\def\n@one{#1}\def\n@two{#2}\mathrel
{\setw@dth{#1}{#2}
\mathop{\hbox to \w@dth{\rightarrowfill}}\limits
\ifx\n@one\empty\else ^{\box\z@}\fi
\ifx\n@two\empty\else _{\box\@ne}\fi}}
\def\t@@^#1{\@ifnextchar_ {\t@^{#1}}{\t@^{#1}_{}}}

\def\t@left^#1_#2{\def\n@one{#1}\def\n@two{#2}\mathrel{\setw@dth{#1}{#2}
\mathop{\hbox to \w@dth{\leftarrowfill}}\limits
\ifx\n@one\empty\else ^{\box\z@}\fi
\ifx\n@two\empty\else _{\box\@ne}\fi}}
\def\t@@left^#1{\@ifnextchar_ {\t@left^{#1}}{\t@left^{#1}_{}}}

\def\two@^#1_#2{\def\n@one{#1}\def\n@two{#2}\mathrel{\setw@dth{#1}{#2}
\mathop{\vcenter{\hbox to \w@dth{\rightarrowfill}\kern-1.7ex
                 \hbox to \w@dth{\rightarrowfill}}%
       }\limits
\ifx\n@one\empty\else ^{\box\z@}\fi
\ifx\n@two\empty\else _{\box\@ne}\fi}}
\def\tw@@^#1{\@ifnextchar_ {\two@^{#1}}{\two@^{#1}_{}}}

\def\tofr@^#1_#2{\def\n@one{#1}\def\n@two{#2}\mathrel{\setw@dth{#1}{#2}
\mathop{\vcenter{\hbox to \w@dth{\rightarrowfill}\kern-1.7ex
                 \hbox to \w@dth{\leftarrowfill}}%
       }\limits
\ifx\n@one\empty\else ^{\box\z@}\fi
\ifx\n@two\empty\else _{\box\@ne}\fi}}
\def\t@fr@^#1{\@ifnextchar_ {\tofr@^{#1}}{\tofr@^{#1}_{}}}

\newdimen\W@dth
\def\setW@dth#1#2{\setbox\z@\hbox{$#1$}\W@dth=\wd\z@
\setbox\@ne\hbox{$#2$}\ifnum\W@dth<\wd\@ne \W@dth=\wd\@ne \fi
\advance\W@dth by 1.2em}

\def\T@^#1_#2{\allowbreak\def\N@one{#1}\def\N@two{#2}\mathrel
{\setW@dth{#1}{#2}
\mathop{\hbox to \W@dth{\rightarrowfill}}\limits
\ifx\N@one\empty\else ^{\box\z@}\fi
\ifx\N@two\empty\else _{\box\@ne}\fi}}
\def\T@@^#1{\@ifnextchar_ {\T@^{#1}}{\T@^{#1}_{}}}

\def\T@left^#1_#2{\def\N@one{#1}\def\N@two{#2}\mathrel{\setW@dth{#1}{#2}
\mathop{\hbox to \W@dth{\leftarrowfill}}\limits
\ifx\N@one\empty\else ^{\box\z@}\fi
\ifx\N@two\empty\else _{\box\@ne}\fi}}
\def\T@@left^#1{\@ifnextchar_ {\T@left^{#1}}{\T@left^{#1}_{}}}

\def\Tofr@^#1_#2{\def\N@one{#1}\def\N@two{#2}\mathrel{\setW@dth{#1}{#2}
\mathop{\vcenter{\hbox to \W@dth{\rightarrowfill}\kern-1.7ex
                 \hbox to \W@dth{\leftarrowfill}}%
       }\limits
\ifx\N@one\empty\else ^{\box\z@}\fi
\ifx\N@two\empty\else _{\box\@ne}\fi}}
\def\T@fr@^#1{\@ifnextchar_ {\Tofr@^{#1}}{\Tofr@^{#1}_{}}}

\def\Two@^#1_#2{\def\N@one{#1}\def\N@two{#2}\mathrel{\setW@dth{#1}{#2}
\mathop{\vcenter{\hbox to \W@dth{\rightarrowfill}\kern-1.7ex
                 \hbox to \W@dth{\rightarrowfill}}%
       }\limits
\ifx\N@one\empty\else ^{\box\z@}\fi
\ifx\N@two\empty\else _{\box\@ne}\fi}}
\def\Tw@@^#1{\@ifnextchar_ {\Two@^{#1}}{\Two@^{#1}_{}}}

\def\to{\@ifnextchar^ {\t@@}{\t@@^{}}}
\def\from{\@ifnextchar^ {\t@@left}{\t@@left^{}}}
\def\two{\@ifnextchar^ {\tw@@}{\tw@@^{}}}
\def\tofro{\@ifnextchar^ {\t@fr@}{\t@fr@^{}}}
\def\To{\@ifnextchar^ {\T@@}{\T@@^{}}}
\def\From{\@ifnextchar^ {\T@@left}{\T@@left^{}}}
\def\Two{\@ifnextchar^ {\Tw@@}{\Tw@@^{}}}
\def\Tofro{\@ifnextchar^ {\T@fr@}{\T@fr@^{}}}

\makeatother

\newcommand{\ropen}[1]{[#1)} 
\newcommand{\lopen}[1]{(#1]} 

\theoremstyle{plain} 

\begin{document}

\title[Polarization under Confirmation Bias]{A Formal Model for Polarization \texorpdfstring{\\}{} under Confirmation Bias in Social Networks}
\titlecomment{}
\thanks{M\'{a}rio S.\ Alvim and Bernardo Amorim were partially supported by CNPq, CAPES, and FAPEMIG\@.
Frank Valencia was partially supported by the ECOS-NORD project FACTS (C19M03) and  the Minciencias project PROMUEVA, BPIN 2021000100160.
}	

\author[M.~S.~Alvim]{M\'{a}rio S.\ Alvim}[a]
\author[B.~Amorim]{Bernardo Amorim}[a]
\author[S.~Knight]{Sophia Knight}[b]
\author[S.~Quintero]{\texorpdfstring{\\}{} Santiago Quintero}[c]
\author[F.~Valencia]{Frank Valencia}[c,d]

\address{Department of Computer Science, UFMG, Brazil}	

\address{Department of Computer Science, University of Minnesota Duluth, USA}	

\address{Laboratoire d'informatique, \'{E}cole Polytechnique, CNRS, France}	

\address{ Pontificia Universidad Javeriana Cali, Colombia}	





\begin{abstract}

    \noindent We describe a model for polarization in multi-agent systems based on Esteban and Ray's standard \review{family of polarization measures from economics}. Agents evolve by updating their beliefs (opinions) based on an underlying influence graph, as in the standard DeGroot model for social learning, but under a \emph{confirmation bias}; i.e., a discounting of opinions of agents with dissimilar views.
    We show that even under this bias polarization eventually vanishes (converges to zero) if the influence graph is strongly-connected. If the influence graph is a regular symmetric circulation, we determine the unique belief value to which all agents converge. Our more insightful result establishes that, under some natural assumptions, if polarization does not eventually vanish then either there is a disconnected subgroup of agents, or some agent influences others more than she is influenced.
    We also prove that polarization does not necessarily vanish in weakly-connected graphs under confirmation bias. Furthermore, we show how our model relates to the classic DeGroot model for social learning.
    We illustrate our model with several simulations of a running example about polarization over vaccines and of other case studies. The theoretical results and simulations will provide insight into the phenomenon of polarization.
\end{abstract}

\maketitle

\section{Introduction}%
\label{sec:introduction}

Distributed systems have changed significantly with the advent of social networks.  In the
previous incarnation of distributed computing~\cite{Lynch96}, the
emphasis was on consistency, fault tolerance, resource management, and
related topics; these were all characterized by \emph{interaction between
processes}. The new era of distributed systems adds an emphasis on the flow of epistemic information (facts, beliefs, lies) and its impact on  democracy and on society at large.

Social networks may facilitate civil discourse by \review{enabling} a prompt exchange of facts, beliefs and opinions among members of a community. Nevertheless,  users in social networks may shape their beliefs by attributing more value to the opinions of influential figures. This common  cognitive bias is known as \emph{authority bias}~\cite{Ramos:19:Book}. Furthermore, social networks often target their users with information \review{that they may already agree with} to keep engagement.  It is known that users tend to give more value to opinions that confirm their own preexisting beliefs~\cite{Aronson10} in another common cognitive bias known as \emph{confirmation bias}.  As a result, social networks can cause their users to become radical and isolated in their own ideological circle, \review{potentially leading to} dangerous splits in society~\cite{Bozdag13} in a phenomenon known as \emph{polarization}~\cite{Aronson10}.

  Indeed, social media platforms have played a key role in the polarization of political processes. Referenda such as Brexit and the Colombian Peace Agreement, as well as recent presidential elections in Brazil and USA are compelling examples of this phenomenon~\cite{Kirby17}. These cases illustrate that messages in social media with elements of extremist ideology in political and public discourse may cause polarization and negatively influence fundamental decision-making processes.

   Consequently, we believe that developing  a model that focuses on central aspects in social networks, such as influence graphs and evolution of users' beliefs, represents a significant contribution to the understanding of the  phenomenon of polarization. In fact, there is a growing interest in the development of models for the analysis of polarization and social influence in networks~\cite{li,proskurnikov,sirbu,gargiulo,alexis,Guerra,myp,degroot,naive,zoe,fblogic,facebook,hunter}. Since polarization involves non-terminating systems with \emph{multiple agents} simultaneously exchanging information (opinions), concurrency models are a natural choice to capture the dynamics of this phenomenon.

\subsubsection*{Our approach}
In this paper we present a multi-agent model for polarization inspired by linear-time models of concurrency where the state of the system evolves in discrete time units (in particular~\cite{tcc,ntcc,Valencia01}). In each time unit, the users, called \emph{agents}, \emph{update} their beliefs about \review{a given} proposition of interest \review{by} taking into account the beliefs of their neighbors \review{through} an underlying weighted \emph{influence graph}. The belief update gives more value to the opinion of agents with higher influence (\emph{authority bias}) and to the opinion of agents with similar views (\emph{confirmation bias}). Furthermore, the model is equipped with a \emph{polarization measure} based on  the seminal work in economics by Esteban and Ray~\cite{Esteban:94:Econometrica}. Polarization is measured at each time unit and it is zero if all agents' beliefs fall within an interval of agreement about the proposition.


Our goal is to explore how the combination of influence graphs and cognitive biases in our model can lead to polarization.  The closest related work is that on DeGroot models~\cite{degroot}. These are the standard linear models for social learning whose analyses can be carried out by standard linear techniques from Markov chains. Nevertheless, a novelty in our model is that its update function extends the classical update  from DeGroot models with confirmation bias. As we elaborate in Section~\ref{sec:degroot}, the extension makes the model no longer linear and thus mathematical tools like Markov chains are not  applicable in a straightforward way. Our model also incorporates a polarization measure in a model for social learning and extends the classical convergence results of DeGroot models to the confirmation bias case.

\subsubsection*{Main Contributions}

We introduce a variant of the standard DeGroot model where agents can update their beliefs under confirmation bias.
By employing techniques from calculus, graph theory, and flow networks, we identify how networks and beliefs are structured, for agents subject to confirmation bias, when polarization \emph{does not} disappear. Furthermore, we illustrate and discuss some perhaps unexpected
aspects of the temporal evolution of polarization by means of a series of
\review{elucidating}
simulations. In particular, we address the non-monotonic evolution of polarization as well as the effect on polarization of various update functions, influence graphs, and initial belief configurations among agents.

The following are our main theoretical contributions. Assuming confirmation bias and some natural conditions about the initial belief values, we show that:
\begin{enumerate}

\item Polarization eventually disappears  (converges to zero) if the influence graph is \emph{strongly-connected} (Definition~\ref{def:strongly-connected}).

    \item If polarization does not disappear then either there is a disconnected subgroup of agents (i.e., the influence graph is not \emph{weakly connected}, see  Definition~\ref{def:weakly-connected}), or some agent influences others more than she is influenced, or all the agents are initially radicalized (i.e., each individual holds the most extreme value either \review{in favor or against the given proposition of interest}).

\item If the influence graph is a regular symmetric circulation (Section~\ref{circulation:section}) we determine the unique belief value all agents converge to.
\end{enumerate}

\noindent
An implementation in Python of the model  and the corresponding simulations presented in this paper are publicly available on GitHub~\cite{website:github-repo}.

 All in all, our formal model, theoretical results, and experimental observations provide insight into the phenomenon of polarization, and are a step toward the design of robust computational models and simulation software for human cognitive and social processes.

\subsubsection*{Organization} 
In Section˜\ref{sec:model} we introduce our formal model, and in Section~\ref{sec:simulations} we provide a series of examples and simulations uncovering interesting new insights and complex characteristics of the evolution of beliefs and polarization under confirmation bias. The first contribution listed above \review{appears} in Section 4 while the other two appear in  Section~\ref{sec:specific-cases}. We discuss DeGroot and other related work in Section~\ref{sec:related-work}, and conclude in Section~\ref{sec:conclusion}. For the sake of readability, \review{the proofs follow in Appendix~\ref{sec:proofs}}.





\section{The Model}%
\label{sec:model}
Here we
present our polarization model, which is
composed of \qm{static} and \qm{dynamic} elements.
We presuppose basic knowledge of calculus and graph theory~\cite{Sohrab:14,Diestel:17}.

\subsection{Static Elements of the Model}%
\label{sec:model-static}

\emph{Static elements} of the model represent a snapshot of a social network
at a given point in time. They include the following components:

\begin{itemize}
 \item A (finite) set $\Agents = \{\agent{0}, \agent{1}, \ldots, \agent{n{-}1} \}$ of $n  \geq 1$ \emph{agents}.

 \item \review{A proposition $p$ representing a declarative sentence, proposing something as being true. We shall refer to $p$ as a \emph{statement} or \emph{proposition}. For example $p$ could be the statement ``\emph{vaccines are safe}'', ``\emph{Brexit was a mistake}'', or ``\emph{climate change is real and is caused by human activity}''.  We shall see next how each agent in $\Agents$  assigns a value to $p$. The sentence $p$ is \emph{atomic} in the sense that the value assigned to $p$ is obtained from $p$ as a whole; i.e., it is not obtained by composing  values assigned to other sentences.
}

 \item A \emph{belief configuration}
    $\Blf:\Agents\rightarrow[0,1]$ \review{such that} for each agent $\agent{i}\in\Agents$, the value $\Blf_{\agent{i}}=\Blf(\agent{i})$ represents the confidence of agent
    $\agent{i}\in\Agents$ in the veracity of proposition $p$. \review{The higher the value $\Blf_{\agent{i}}$, the higher the confidence of agent $\agent{i}$ in the veracity of $p$.}
    Extreme values $0$ and $1$ represent a firm belief of agent $\agent{i}$ in, respectively, the falsehood or truth of  $p$. \review{A belief configuration $\Blf$ can also be represented as a tuple $(\Blf_{\agent{0}},\ldots, \Blf_{\agent{n-1}})$. Given the set  of agents $\Agents$, we  use $[0,1]^\Agents$ to denote the set all belief configurations over $\Agents$.}

 \item A \emph{polarization measure} $\Pol:[0,1]^\Agents\rightarrow\R^+$
 mapping belief configurations to the non-negative real numbers. \review{Given a belief configuration  $B=(\Blf_{\agent{0}},\ldots, \Blf_{\agent{n-1}})$, the value $\Pfun{\Blf}$ indicates the polarization among all the agents in $\Agents$ given their beliefs $\Blf_{\agent{0}},\ldots, \Blf_{\agent{n-1}}$ about the veracity of the statement $p$. The higher the value $\Pfun{\Blf}$, the higher the polarization amongst the agents in $\Agents$.}
\end{itemize}

\noindent
There are several polarization measures described in the literature.
In this work we  \review{employ the influential family of measures} proposed by
Esteban and Ray~\cite{Esteban:94:Econometrica}.

\review{In the rest of the paper, we will use the following notion.} We say that  $(\pi, y)=(\pi_0, \allowbreak \pi_1, \allowbreak \ldots, \allowbreak \pi_{k{-}1},  \allowbreak y_0, \allowbreak y_1, \allowbreak \ldots, \allowbreak y_{k{-}1})$ is  a \emph{distribution}  if \review{$y \in \R^{k}$}, $\sum_{i=0}^{k-1}\pi_i = 1$ and for every $i,j$ we have $\pi_i \geq 0$ and $y_i \neq y_j$ whenever $j\neq i$. We use $\mathcal{D}$ to denote the set of all distributions.

\review{
\begin{defi}[Esteban-Ray Polarization,~\review{\cite{Esteban:94:Econometrica}}]%
\label{def:poler} An \emph{Esteban-Ray polarization measure} is a
mapping $\PolER: \mathcal{D} \to \R^+$ for which there are
constants $K > 0$ and $\alpha \in (0,2)$ such that for every $(\pi, y)=(\pi_0, \allowbreak \pi_1, \allowbreak \ldots, \allowbreak \pi_{k{-}1},  \allowbreak y_0, \allowbreak y_1, \allowbreak \ldots, \allowbreak y_{k{-}1}) \in \mathcal{D}$ we have
\begin{equation*}
    \PfunER{\pi, y} = K \sum_{i=0}^{k-1} \sum_{j=0}^{k-1} \pi_i^{1+\alpha} \pi_j | y_i - y_j |.
\end{equation*}
\end{defi}}

The higher the value of $\PfunER{\pi, y}$, the more polarized
distribution $(\pi,y)$ is.
The measure captures the intuition that
polarization is accentuated by both intra-group homogeneity
and inter-group heterogeneity.
Moreover, it assumes that the total polarization
is the sum of the effects of individual agents on one another.
This measure (family) can be derived from a set of intuitively reasonable axioms~\cite{Esteban:94:Econometrica}, \review{which are presented in Appendix~\ref{sec:polar-axioms}.}
\review{Succinctly, the measure considers a society as highly polarized when
agents can be divided into two clusters of similar size, one in which everyone has
a high level of confidence in the veracity of the proposition, and the other
in which everyone has a low level of confidence in the veracity of that same
proposition.
On the other hand, the measure considers a society as not polarized at all
when all individuals share a similar level of belief, and considers it
as slightly polarized when all individuals hold different levels of belief,
without forming distinctive clusters of similar opinions
(i.e., the spread of opinions is diffuse.)}

Note that $\PolER$ is defined on a discrete distribution,
whereas in our model a general polarization metric is defined on
a belief configuration $\Blf:\Agents\rightarrow[0,1]$.
To apply $\PolER$ to our setup \review{in~\cite{Alvim:19:FC} we converted} the belief configuration
$\Blf$ into an appropriate distribution $(\pi,y)$.

First we need some notation: Let $D_k$ be a discretization of the interval $[0,1]$  into
$k > 0$ consecutive non-overlapping, non-empty intervals (\emph{bins})  $I_0,I_1,\ldots, I_{k-1}$. We use the term \emph{borderline  points} of $D_k$ to refer to the end-points  of $I_0,I_1,\ldots, I_{k-1}$  different from 0 and 1. Given a belief configuration $B$, define the \emph{belief distribution of $B$ in $D_k$} as $\BD{B}{D_k}=(\pi_0, \allowbreak \pi_1, \allowbreak \ldots, \allowbreak \pi_{k{-}1},  \allowbreak y_0, \allowbreak y_1, \allowbreak \ldots, \allowbreak y_{k{-}1})$ where each $y_i$ is the mid-point of $I_i$,
and  $\pi_i$ is the fraction of agents having their belief in $I_{i}$.

\begin{defi}[$k$-bin polarization,~\review{\cite{Alvim:19:FC}}]\label{k-bin:def}

\review{An Esteban-Ray polarization measure for belief configurations over  $D_k$
is a mapping $\Pol:[0,1]^\Agents\rightarrow\R^+$  such that for some Esteban-Ray polarization measure $\PolER$, we have
\[ \Pfun{\Blf} = \PfunER{\BD{B}{D_k}} \]
for every belief configuration $B \in [0,1]^\Agents. $}
\end{defi}

Notice that when there is consensus about the proposition $p$ of interest,
i.e., when all agents in belief configuration $\Blf$ hold the same belief
value, we have $\Pfun{\Blf}=0$.
This happens exactly when all agents' beliefs fall within the same bin of the underlying discretization $D_k$. The following property is an easy consequence from Definition~\ref{def:poler} and Definition~\ref{k-bin:def}.

\begin{restatable}[Zero Polarization]{prop}{respolconsensus}%
\label{pol-consensus} Let $\Pol$ be a Esteban-Ray polarization measure for belief configurations over a discretization $D_k=I_0,\ldots,I_{k-1}$. Then
 $\Pfun{\Blf}=0$ iff there exists $m\in\{0,\ldots,k{-}1\}$ \review{such that} for all $i\in\Agents$,
 $\Blf_{\agent{i}} \in I_m$.
\end{restatable}

\subsection{Dynamic Elements of the Model}%
\label{sec:model-dynamic}

\emph{Dynamic elements} formalize the evolution of agents' beliefs as they interact over time and are exposed to different opinions. They include:

\begin{itemize}
\item A \emph{time frame}
\[\calt{ = }\{0, 1, 2, \ldots \}\]
representing the discrete passage of time.
\item A \emph{family of belief configurations}
\[\{\Blft{t} : \Agents \rightarrow [0,1]\}_{t\in{\calt}}\]
\review{such that}\ each $\Blft{t}$ is the belief configuration of agents in
$\Agents$ \review{with respect to} proposition $p$ at time step $t\in\calt$.

\item A \emph{weighted directed graph}
\[\Inter : \Agents \times \Agents  \rightarrow [0,1].\]
The value $\Inter(\agent{i},\agent{j})$, written $\Ifun{\agent{i}}{\agent{j}}$,
represents the \emph{direct influence} that agent $i$ has \review{on} agent $j$, or the \emph{weight} $\agent{i}$ carries with $\agent{j}$.
A higher value means \review{that agent $j$ will have higher confidence in agent $i$'s opinion and, therefore, will give this opinion more weight when incorporating it into its own.} Conversely,  $\Inter_{\agent{i},\agent{j}}$ can also be viewed as the \emph{trust} or \emph{confidence} that  $j$ has \review{in} $i$.  We assume that $\Ifun{\agent{i}}{\agent{i}}=1$ \review{for every agent $i$}, meaning that agents are self-confident. We shall often refer to $\Inter$ simply as the \emph{influence} (graph) $\Inter$.

We distinguish, however, the direct influence $\Inter_{\agent{i},\agent{j}}$ that $\agent{i}$ has \review{on}  $\agent{j}$
from the \textit{overall effect} of $\agent{i}$ \review{on} $\agent{j}$'s belief.
This effect is a combination of various factors, including
direct influence, their current opinions, the topology of the influence graph, and how agents reason. This overall effect is captured by the update function below.

\item  An \emph{update function}
\[\Upd : (\Blft{t},\Inter) \mapsto \Blft{t+1}\]
mapping belief configuration $\Blft{t}$ at time $t$ and influence graph $\Inter$ to new belief configuration $\Blft{t+1}$ at time $t + 1$.
This function models the evolution of agents' beliefs over time.
We adopt the following premises.
\end{itemize}

\begin{enumerate}[(i)]
    \item \emph{Agents present some Bayesian reasoning}:
    \review{Agents'} beliefs are updated \review{at} every time step by combining their current belief with a \emph{correction term} that incorporates the new evidence they are exposed to
    in that step --i.e., other agents' opinions.
    More precisely, when agent $\agent{j}$ interacts with agent $\agent{i}$,
    the former affects the latter moving $\agent{i}$'s belief towards $\agent{j}$'s,
    proportionally to the difference $\Bfun{\agent{j}}{t}{ - }\Bfun{\agent{i}}{t}$ in their beliefs.
    The intensity of the move is proportional to the influence $\Ifun{\agent{j}}{\agent{i}}$
    that $\agent{j}$ carries with $\agent{i}$.
    The update function produces an overall correction term for each agent as the average of all
    other agents' effects on that agent,  and then incorporates this term into the agent's current belief.~\footnote{\ Note that this assumption implies that an agent has, in effect, an influence on \review{itself},
    and hence cannot be used as a ``puppet'' who immediately assumes another's agent's belief.}
    The factor $\Ifun{\agent{j}}{\agent{i}}$  allows the model to capture
    \emph{authority bias}~\cite{Ramos:19:Book},
    by which agents' influences on each other may have different intensities (by, e.g., giving
    higher weight to an authority's opinion).

    \item \emph{Agents may be prone to confirmation bias}:
    Agents may give more weight to evidence supporting their
    current beliefs while discounting evidence contradicting them,
    independently from its source.
    This behavior in known in the psychology literature as
    \emph{confirmation bias}~\cite{Aronson10},
    and is captured in our model as follows.

    When agent $\agent{j}$ interacts with agent $\agent{i}$ at time $t$, the update function moves agent $\agent{i}$'s belief toward
    that of agent $\agent{j}$, proportionally to the influence $\Ifun{\agent{j}}{\agent{i}}$ of $\agent{j}$ on $\agent{i}$, but with a caveat: the move is stronger when $\agent{j}$’s belief is similar to $\agent{i}$’s than when it is dissimilar. \review{This is realized below by making the move proportional to what we shall call \emph{confirmation-bias factor}  $\CBfun{i}{j}{t} = 1- |\Bfun{j}{t}{-}\Bfun{i}{t}|$. Clearly, the closer the beliefs of agents $i$ and $j$ at time $t$, the higher the factor $\CBfun{i}{j}{t}$.}
\end{enumerate}

\noindent
The premises above are formally captured in the following update-function. As usual, given a set $S$, we shall use $|S|$ to denote the cardinality of $S$.

\begin{defi}[Confirmation-bias \review{update function}]%
\label{def:confirmation-bias}
    Let $\Blft{t}$ be a belief configuration at time $t\in\calt$, and  $\Inter$ be an influence graph. The \emph{confirmation-bias update-function} is the map  $\UpdCB:({\Blft{t}},{\Inter})\mapsto\Blft{t+1}$ with  $\Blft{t+1}$ given by
    \begin{equation}\label{eq:confirmation-bias}
        \Bfun{\agent{i}}{t+1} = \Bfun{\agent{i}}{t} + \frac{1}{|\Agents_{\agent{i}}|} \sum_{\agent{j} \in \Agents_{\agent{i}}}
        \CBfun{i}{j}{t} \, \Ifun{j}{i} \, (\Bfun{j}{t} - \Bfun{i}{t}),
   \end{equation}
   for every agent $\agent{i}\in\Agents$,
   where $\Agents_{\agent{i}}=\{\agent{j}\in\Agents \mid \Inter_{j,i} > 0 \}$ is the set of \emph{neighbors} of $\agent{i}$ and  $\CBfun{i}{j}{t} = 1- |\Bfun{j}{t}{-}\Bfun{i}{t}|$ is the \emph{confirmation-bias factor} of  $i$ \review{with respect to} $j$ given their beliefs at time $t$.

\end{defi}

The expression $\nicefrac{1}{|\Agents_{\agent{i}}|} \sum_{\agent{j} \in \Agents_{\agent{i}}} \CBfun{i}{j}{t} \, \Ifun{j}{i} \, (\Bfun{j}{t} - \Bfun{i}{t})$ on the right-hand side of Definition~\ref{def:confirmation-bias}
    is a \emph{correction term} incorporated into agent $\agent{i}$'s
    original belief $\Bfun{\agent{i}}{t}$ at time $t$.
    The correction is the average of the effect of each
    neighbor $\agent{j}\in\Agents_{i}$ on agent $\agent{i}$'s belief
    at that time step.
    The value  $\Bfun{\agent{i}}{t+1}$ is the resulting updated belief of
    agent $\agent{i}$ at time $t + 1$.
    \review{By rewriting~\eqref{eq:confirmation-bias} as
    $\Bfun{\agent{i}}{t+1} = \nicefrac{1}{|\Agents_{\agent{i}}|} \sum_{\agent{j} \in \Agents_{\agent{i}}}
    \CBfun{i}{j}{t} \, \Ifun{j}{i} \, \Bfun{j}{t} +  \left(1 - \CBfun{i}{j}{t} \, \Ifun{j}{i}\right) \Bfun{\agent{i}}{t}$,
    it is easy to verify that $\Bfun{\agent{i}}{t+1} \in [0,1]$,
    since: (i) we divide the result of the summation by the number of terms; and  (ii) each term of the summation also belongs to the interval $[0, 1]$, as it is a convex combination of the beliefs of agent $i$'s neighbors at time $t$ and its own belief, both in the interval $[0,1]$.}

The confirmation-bias factor $\CBfun{i}{j}{t}$
lies in the interval $[0,1]$, and the lower its value, the more agent $\agent{i}$ discounts
the opinion provided by agent $\agent{j}$ when incorporating it.
It is maximum when agents' beliefs are identical, and minimum
\review{when their beliefs are} extreme opposites.

\begin{rem}[Classical Update: Authority Non-Confirmatory Bias]\label{rem:auto-bias}
In this paper we focus on confirmation-bias update and, unless otherwise
stated, assume the underlying function is given by Definition~\ref{def:confirmation-bias}.
Nevertheless, in Sections~\ref{circulation:section} and~\ref{sec:degroot}
we will consider a \emph{classical update} $\UpdR:({\Blft{t}},{\Inter}){\mapsto}\Blft{t+1}$ that captures
non-confirmatory authority-bias and
is obtained by replacing the confirmation-bias factor $\CBfun{i}{j}{t}$
in Definition~\ref{def:confirmation-bias} with 1.
That is,
\begin{equation}\label{eq:auto-bias}
\Bfun{\agent{i}}{t+1} = \Bfun{\agent{i}}{t} + \frac{1}{|\Agents_{\agent{i}}|} \sum_{\agent{j} \in \Agents_{\agent{i}}}
         \Ifun{j}{i} \, (\Bfun{j}{t}{-}\Bfun{i}{t}). \end{equation}
We refer to this function as \emph{classical} because it is
closely related to the standard update function of the DeGroot
models for social learning from Economics~\cite{degroot}. This correspondence will be formalized in Section~\ref{sec:degroot}.
\end{rem}

\begin{rem}\review{A preliminary and slightly different version of the biases in Definition~\ref{def:confirmation-bias} and Remark~\ref{rem:auto-bias} using in Equations~\ref{eq:confirmation-bias} and~\ref{eq:auto-bias} the set $\Agents$ instead of $\Agents_{\agent{i}}$  were given~\cite{Alvim:19:FC}. As a consequence these preliminary definitions take into account the weighted average of all agents' beliefs rather that only those of the agents that have an influence over agent $\agent{i}$.} \end{rem}


\section{Running Example and Simulations}%
\label{sec:simulations}
In this section we present a running example, as well as several simulations, that motivate our theoretical results from the following sections. We start by stating some assumptions that will be adopted throughout this section.

\subsection{General assumptions}

Recall that we assume  $\Ifun{\agent{i}}{\agent{i}}=1$ for every $i\in\Agents$.
However, for simplicity, in all figures of influence graphs we omit self-loops.
In all cases we limit our analyses to a fixed number of time steps.
We compute a polarization measure from Definition~\ref{k-bin:def}
with parameters $\alpha = 1.6$,
as suggested by Esteban and Ray~\cite{Esteban:94:Econometrica},
and \review{$K = 1,000$}.
Moreover, we employ a discretization $D_k$ of the interval $[0,1]$ into $k = 5$ bins,
each representing a possible general position
\review{with respect to} the veracity of the proposition $p$ of interest:
\begin{itemize}
    \item \emph{strongly against}: $\ropen{0,0.20}$;
    \item \emph{fairly against}: $\ropen{0.20,0.40}$;
    \item \emph{neutral/unsure}: $\ropen{0.40,0.60}$;
    \item \emph{fairly in favour}: $\ropen{0.60,0.80}$; and
    \item \emph{strongly in favour}: $[0.80,1]$.\footnote{\ Recall from Definition~\ref{k-bin:def} that our model allows arbitrary discretizations $D_k$ --i.e., different number of bins, with not-necessarily
uniform widths-- depending on the scenario of interest.}
\end{itemize}
In all definitions we let $\Agents = \{0, 1,\ldots, n{-}1 \}$,
and $\agent{i},\agent{j} \in \Agents$ be generic agents.

\subsection{Running example}
As a motivating example we consider the following hypothetical situation.

\begin{exa}[Vaccine Polarization]%
\label{running-example}
Consider the sentence ``vaccines are safe'' as the proposition $p$ of interest.
Assume a set $\Agents$ of $6$ agents that is initially \emph{extremely polarized} about $p$:
Agents 0 and 5 are absolutely confident, respectively, in the falsehood or
truth of $p$, whereas the others are equally split into strongly in favour and strongly against $p$.
Consider first the situation described by the influence graph in Figure~\ref{fig:circulation-graph-a}.
Nodes 0, 1 and 2 represent anti-vaxxers, whereas the rest are pro-vaxxers.
In particular, note that although initially in total disagreement about $p$, Agent 5 carries a lot of weight with Agent 0.
In contrast,  Agent $0$'s opinion is very close to that of Agents 1 and 2, even if they do not have any direct influence over him.
Hence the evolution of Agent $0$'s beliefs will be mostly shaped by that of Agent $5$.
\review{As can be observed in the evolution of agents' opinions
in Figure~\ref{fig:circulation-graph-a},
Agent 0 (represented by the purple line) moves from being
initially strongly against $p$
(i.e., having an opinion in the range of $\ropen{0.00,0.20}$ at time step $0$)
to being fairly in favour of $p$
(i.e., having an opinion in the range of $\ropen{0.60,0.80}$)
around time step 8.}
Moreover, polarization eventually vanishes (i.e., becomes zero) around time 20, as agents reach the consensus of being fairly against  $p$.

\review{
\begin{figure}[tbp]
    \centering
    \begin{subfigure}[m]{1.0\textwidth}
      \centering
      \raisebox{0.10\height}{\includegraphics[width=0.45\textwidth]{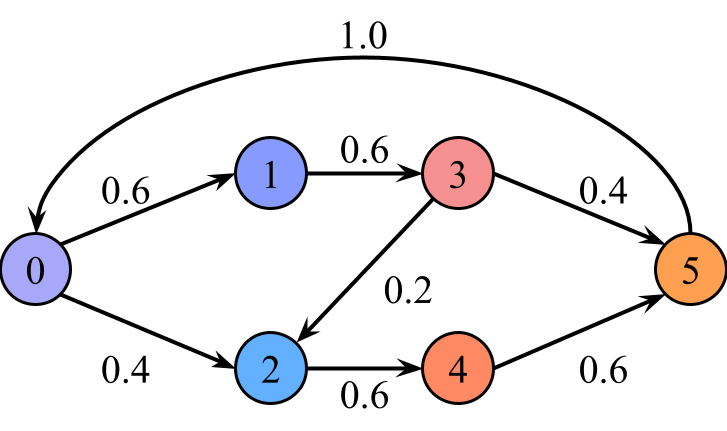}}
      \hfill
      \includegraphics[width=0.45\textwidth]{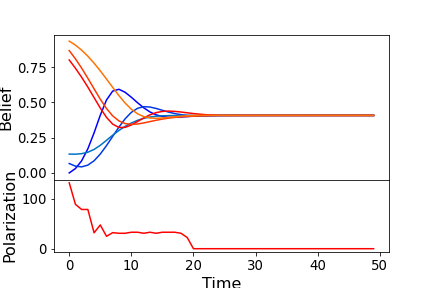}
      \caption{\review{Influence graph $\Inter$ and the corresponding evolution of beliefs and polarization for Example~\ref{running-example}.}}%
      \label{fig:circulation-graph-a}
    \end{subfigure}
    \\
    \begin{subfigure}[m]{1.0\textwidth}
      \centering
      \raisebox{0.10\height}{\includegraphics[width=0.45\textwidth]{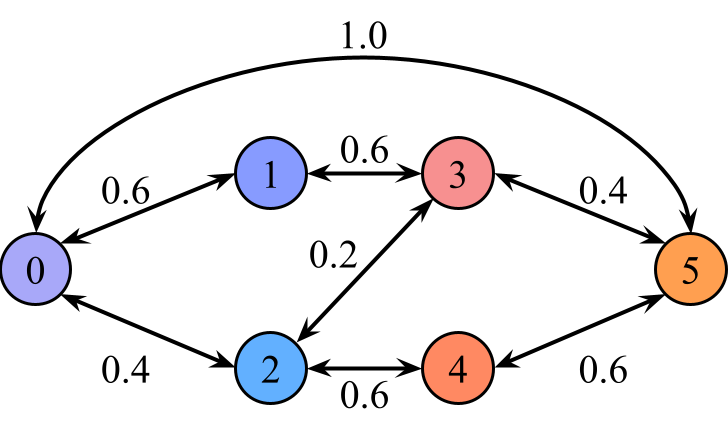}}
      \hfill
      \includegraphics[width=0.45\textwidth]{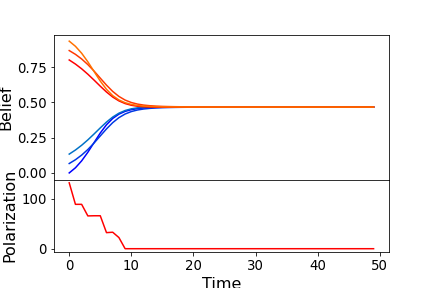}
      \caption{\review{Adding inverse influences to Figure~\ref{fig:circulation-graph-a}.}}%
      \label{fig:circulation-graph-c}
    \end{subfigure}
    \\
    \begin{subfigure}[m]{1.0\textwidth}
      \centering
      \raisebox{0.10\height}{\includegraphics[width=0.45\textwidth]{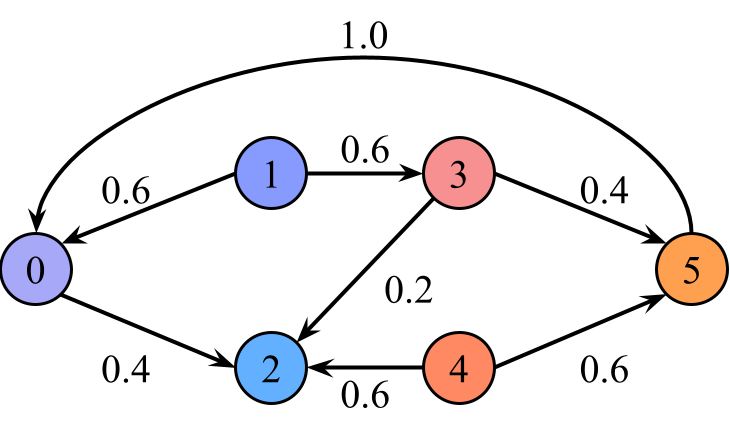}}
      \hfill
      \includegraphics[width=0.45\textwidth]{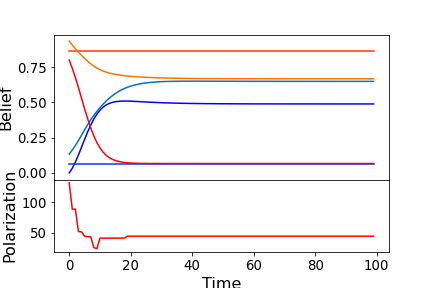}
      \caption{\review{Inversion of the influences $\Ifun{0}{1}$ and $\Ifun{2}{4}$ in Figure~\ref{fig:circulation-graph-a}.}}%
      \label{fig:weakly-graph}
    \end{subfigure}
    \caption{\review{Influence graphs and the corresponding evolution of beliefs
    and polarization for Example~\ref{running-example}.
    In each figure, the left-hand size is a graph representation of
    an influence graph, with nodes representing agents
    and each arrow from an agent $i$ to an agent $j$ labelled with
    the influence $\Ifun{\agent{i}}{\agent{j}}$ the former carries on
    the latter.
    In each graph on the right-hand side, the $x$-axis represents the passage
    of time, and the $y$-axis is divided into:
    (i) an upper half depicting the evolution of beliefs for each agent at each time step
    (with each line representing the agent of same color in the corresponding influence graph);
    and
    (ii) a lower half depicting the corresponding measure of polarization of the
    network at each time step.
    In all cases the initial belief values for Agents $\agent{0}$ to $\agent{5}$ are, respectively, $0.0$, $0.0\overline{6}$, $0.1\overline{3}$, $0.8$, $0.8\overline{6}$, and $0.9\overline{3}$.}
    }%
    \label{running-example:figure}
\end{figure}
}

Now consider the influence graph in Figure~\ref{fig:circulation-graph-c}, which is similar to Figure~\ref{fig:circulation-graph-a}, but with reciprocal influences (i.e., the influence of $i$ over $j$ is the same as the influence of $j$ over $i$). Now  Agents 1 and 2  do have direct influences over Agent 0, so the evolution of Agent $0$'s belief will be partly shaped by initially opposed agents: Agent 5 and the anti-vaxxers. But since Agent $0$'s opinion is very close to that of Agents 1 and 2, the confirmation-bias factor will help \review{in keeping}
Agent $0$'s opinion close to their opinion against $p$. In particular, in contrast to the situation in
\review{Figure~\ref{fig:circulation-graph-a},}
Agent $0$ never becomes in favour of $p$. The evolution of the agents' opinions and their polarization is shown in
\review{Figure~\ref{fig:circulation-graph-c}.}
Notice that polarization vanishes around time 8 as the agents reach consensus, but this time they are more positive about (less against) $p$ than in the first situation.

Finally, consider the situation in Figure~\ref{fig:weakly-graph} obtained from Figure~\ref{fig:circulation-graph-a} by inverting the influences of Agent 0 over Agent 1 and
Agent 2 over Agent 4. Notice that Agents 1 and 4 are no longer influenced by anyone though they influence others. Thus, as shown in
\review{Figure~\ref{fig:weakly-graph}, }
their beliefs do not change over time,
which means that the group does not reach consensus and polarization never disappears though it is considerably reduced. \review{\exaend}
\end{exa}
The above example illustrates complex non-monotonic,  overlapping, convergent, and non-convergent evolution of agent beliefs and polarization
even in a small case with $n = 6$ agents.
Next we shall consider richer simulations on a greater variety of scenarios.
These are instrumental in providing insight for theoretical results to be proven in the following sections.

\subsection{Simulations}%
\label{sec:simulations-description}

Here we  present simulations for several influence graph topologies
with \review{$n = 1,000$} agents (unless stated otherwise), which illustrate more complex behavior
emerging from con\-fir\-ma\-tion-bias interaction among agents.
Our theoretical results in the next sections bring insight into
the evolution of beliefs and polarization depending on graph topologies.

Next we provide the contexts used in our simulations.

\subsubsection*{Initial belief configurations}
We consider the following initial belief configurations, depicted in Figure~\ref{fig:initial-belief-configurations}:
\begin{figure}[tbp]
    \centering
    \includegraphics[width=\linewidth]{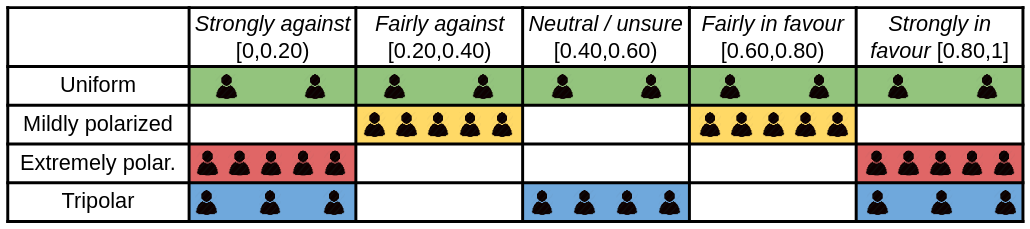}
    \caption{\review{Depiction of the general shape of initial belief configurations (formally defined in Section~\ref{sec:simulations-description}),  recreated here for a small group of only $n = 10$ agents for exemplification purposes (since in the simulations the number of agents used is much larger).
    Each row represents an initial belief configuration, and the 10
    agents' opinions are distributed into columns according to their
    level of belief in the veracity of the proposition of interest.
    Empty cells represent positions that are not held by any agent,
    whereas the other cells have agents with beliefs uniformly
    distributed in the corresponding range (as formally
    defined in Section~\ref{sec:simulations-description}).
    Colors are used only to facilitate the visualization of each configuration.
    }
    }%
    \label{fig:initial-belief-configurations}
\end{figure}


\begin{itemize}
    \item A \emph{uniform} belief configuration representing a set of agents whose beliefs are as varied as possible, all equally spaced in the interval $[0, 1]$\review{: for every $i$,
    \begin{align*}
        \Bfun{i}{0} = \nicefrac{i}{(n{-}1)}.
    \end{align*}
    }

    \item A \emph{mildly polarized} belief configuration with agents evenly split into two groups with moderately dissimilar inter-group beliefs compared to intra-group beliefs\review{: for every $i$,
    \begin{align*}
        \Bfun{i}{0} =
        \begin{cases}
            \nicefrac{0.2 i}{\ceil{\nicefrac{n}{2}}} + 0.2, & \text{if $i<\ceil{\nicefrac{n}{2}}$,} \\
            \nicefrac{0.2 (i{-}\ceil{\nicefrac{n}{2}})}{(n{-}\ceil{\nicefrac{n}{2}})} + 0.6 & \text{otherwise.}
        \end{cases}
    \end{align*}
    }

    \item An \emph{extremely polarized} belief configuration
    representing a situation in which half of the
    agents strongly believe the proposition, whereas
    half strongly disbelieve it\review{: for every $i$,
    \begin{align*}
        \Bfun{i}{0} =
        \begin{cases}
            \nicefrac{0.2 i}{\ceil{\nicefrac{n}{2}}}, & \text{if $i<\ceil{\nicefrac{n}{2}}$,} \\
            \nicefrac{0.2 (i{-}\ceil{\nicefrac{n}{2}})}{(n{-}\ceil{\nicefrac{n}{2}})} + 0.8, & \text{otherwise.}
        \end{cases}
    \end{align*}
    }

    \item A \emph{tripolar} configuration with
    agents divided into three
    groups\review{: for every $i$,
    \begin{align*}
        \Bfun{i}{0} =
        \begin{cases}
            \nicefrac{0.2 i}{\lfloor\nicefrac{n}{3}\rfloor}, & \text{if $i < \lfloor{\nicefrac{n}{3}}\rfloor$,} \\
            \nicefrac{0.2 (i{-}\lfloor{\nicefrac{n}{3}}\rfloor)}{(\ceil{\nicefrac{2n}{3}}{-}\lfloor{\nicefrac{n}{3}}\rfloor)} + 0.4, & \text{if $\lfloor{\nicefrac{n}{3}}\rfloor \leq i < \ceil{\nicefrac{2n}{3}}$,} \\
            \nicefrac{0.2 (i{-}\ceil{\nicefrac{2n}{3}})}{(n{-}\ceil{\nicefrac{2n}{3}})} + 0.8, & \text{otherwise.}
        \end{cases}
    \end{align*}
    }
\end{itemize}

\subsubsection*{Influence graphs}
As for influence graphs, we consider the following ones, depicted in Figure~\ref{fig:interaction-graphs}:

\begin{figure}[tbp]
    \centering
    \begin{subfigure}[t]{0.34\textwidth}
      \centering
      \includegraphics[width=0.5\textwidth]{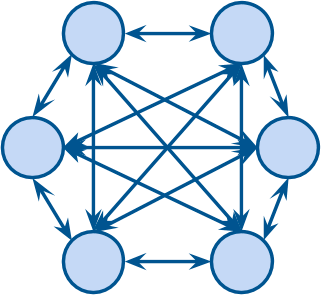}
      \caption{$C$-clique.}%
      \label{fig:interaction-graphs-clique}
    \end{subfigure}
    \hfill
    \begin{subfigure}[t]{0.32\textwidth}
      \centering
      \includegraphics[width=0.53125\textwidth]{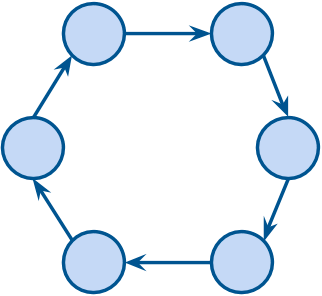}
      \caption{Circular.}%
      \label{fig:interaction-graphs-circular}
    \end{subfigure}
    \hfill
    \begin{subfigure}[t]{0.32\textwidth}
      \centering
      \includegraphics[width=.85\textwidth]{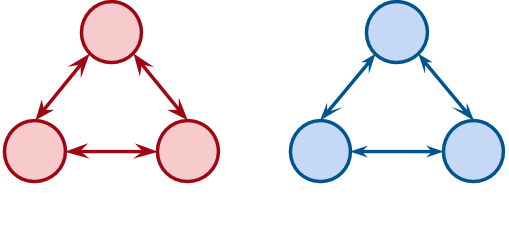}
      \caption{Disconnected groups.}%
      \label{fig:interaction-graphs-disconnected}
    \end{subfigure}
    \\
    \vspace{2mm}
    \begin{subfigure}[t]{0.34\textwidth}
      \centering
      \includegraphics[width=0.8\textwidth]{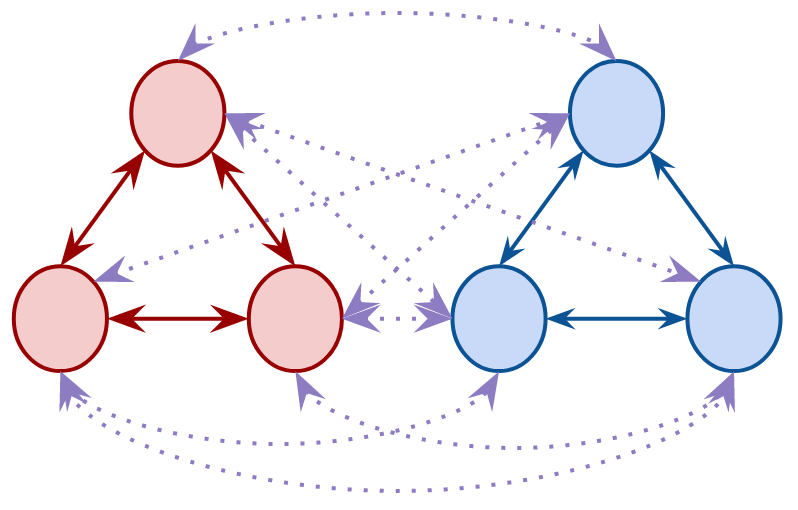}
      \caption{Faintly communicating groups.}%
      \label{fig:interaction-graphs-faintly}
    \end{subfigure}
    \hfill
    \begin{subfigure}[t]{0.32\textwidth}
      \centering
      \includegraphics[width=0.8\textwidth]{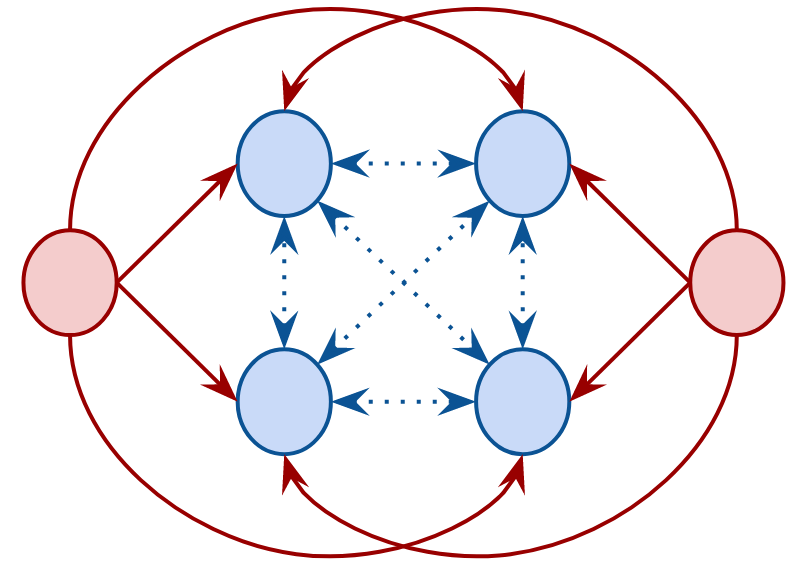}
      \caption{Unrelenting influencers.}%
      \label{fig:interaction-graphs-unrelenting}
    \end{subfigure}
    \hfill
    \begin{subfigure}[t]{0.32\textwidth}
      \centering
      \includegraphics[width=0.8\textwidth]{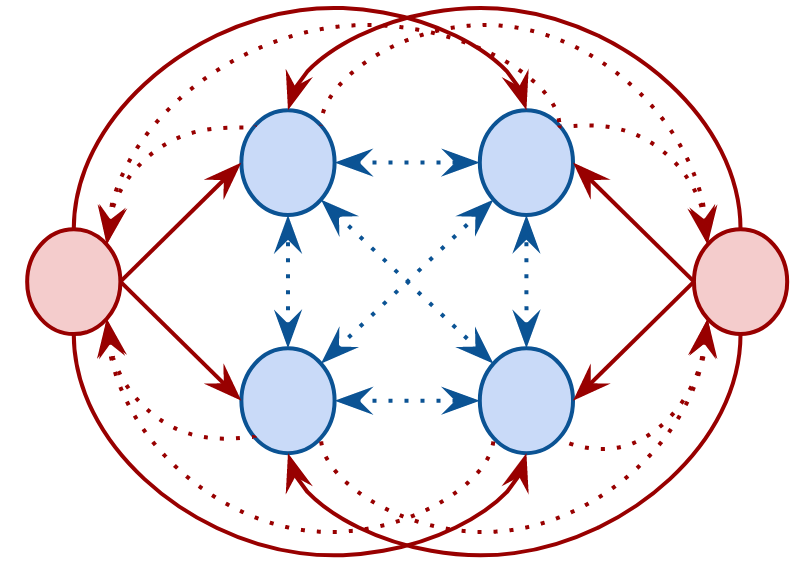}
      \caption{Malleable influencers.}%
      \label{fig:interaction-graphs-malleable}
    \end{subfigure}
    \caption{\review{Depiction of the general shape of influence graphs (formally defined in Section~\ref{sec:simulations-description}),  recreated here for a small group of only $n = 6$ agents for exemplification purposes (since in the simulations the number of agents used is much larger).
    Solid (respectively, dashed) arrows indicate that the originating agent has relatively high (respectively, low) influence on the receiving agent.
    In influence graphs in which agents can be divided into groups with different behaviors, colors are used to indicate such groups.
    }}%
    \label{fig:interaction-graphs}
\end{figure}

\begin{itemize}
    \item A \emph{$C$-clique} influence graph $\Interclique$,
    in which each agent influences every other with constant
    value $C=0.5$\review{: for every $i,j$ such that  $i{\neq}j$,
    \begin{align*}
        \Ifunclique{\agent{i}}{\agent{j}} = 0.5~.
    \end{align*}
    }
    This represents the particular case of a social network in
    which all agents interact among themselves, and are all immune
    to authority bias.

    \item A \emph{circular} influence graph $\Intercircular$
    representing a social network in which agents can be organized
    in a circle in such a way each agent is only influenced by its
    predecessor and only influences its successor\review{: for every $i,j$ such that  $i{\neq}j$,
    \begin{align*}
        \Ifuncircular{\agent{i}}{\agent{j}} =
        \begin{cases}
            0.5, & \text{if (i{+}1)\,\text{mod}\,n = j,} \\
            0, & \text{otherwise.}
        \end{cases}
    \end{align*}
    }
    This is a simple instance of a balanced graph (in which
    each agent's influence on others is as high as the influence received,
    as in Definition~\ref{def:circulation} ahead), which is a pattern commonly
    encountered in some sub-networks.

    \item A \emph{disconnected} influence graph $\Interdisconnected$
    representing a social network sharply divided into two groups in such a way that agents within the same group
    can considerably influence each other, but not at all
    agents in the other group\review{: for every $i,j$ such that  $i{\neq}j$,
    \begin{align*}
        \Ifundisconnected{\agent{i}}{\agent{j}} =
        \begin{cases}
            0.5, & \text{if $\agent{i},\agent{j}$ are both
     ${<} \ceil{\nicefrac{n}{2}}$ or both ${\geq} \ceil{\nicefrac{n}{2}}$,} \\
            0, & \text{otherwise.}
        \end{cases}
    \end{align*}
    }

    \item A \emph{faintly communicating} influence graph $\Interfaint$
    representing a social network divided into two groups that
    evolve mostly separately, with only faint communication between them.
    More precisely, agents from within the same group influence each other
    much more strongly than agents from different groups\review{: for every $i,j$ such that  $i{\neq}j$,
    \begin{align*}
        \Ifunfaint{\agent{i}}{\agent{j}} =
        \begin{cases}
            0.5, & \text{if $\agent{i},\agent{j}$ are both
     ${<} \ceil{\nicefrac{n}{2}}$ or both ${\geq} \ceil{\nicefrac{n}{2}}$,} \\
            0.1, & \text{otherwise.}
        \end{cases}
    \end{align*}
    }
    This \review{could represent} a small, ordinary social network, where some close groups of agents have strong influence on one another, and all agents communicate to some extent.

    \item An \emph{unrelenting influencers} influence graph $\Interunrelenting$
    representing a scenario in which two agents
    (say, $\agent{0}$ and $\agent{n{-}1}$)
    exert significantly stronger influence on every other agent than these other agents have among themselves\review{: for every $i,j$ such that  $i{\neq}j$,
    \begin{align*}
        \Ifununrelenting{\agent{i}}{\agent{j}}=
        \begin{cases}
            0.6, & \text{if $i = 0$ and $j \neq n{-}1$ or $i = n{-}1$
    and $j \neq 0$,} \\
            0, & \text{if $j=0$ or $j=n{-}1$,} \\
            0.1, & \text{if $0\neq i \neq n{-}{1}$ and $0 \neq j \neq n{-}{1}$.}
        \end{cases}
    \end{align*}
    }
    This could represent, e.g., a social network
    in which two totalitarian media companies dominate the news
    market, both with similarly high levels of influence on
    all agents.
    The networks have clear agendas to push forward, and are not influenced in a meaningful way by other agents.

    \item A \emph{malleable influencers} influence graph $\Intermalleable$
    representing a social network in which two agents have
    strong influence on every other agent,
    but are barely influenced by anyone else\review{: for every $i,j$ such that  $i{\neq}j$,
    \begin{align*}
        \Ifunmalleable{\agent{i}}{\agent{j}}=
        \begin{cases}
            0.8, & \text{if $i = 0$ and $j \neq n{-}1$,} \\
            0.4, & \text{if $i = n{-}1$ and $j \neq 0$,} \\
            0.1, & \text{if $j=0$ or $j=n{-}1$,} \\
            0.1, & \text{if $0\neq i \neq n{-}{1}$ and $0 \neq j \neq n{-}{1}$.}
        \end{cases}
    \end{align*}
    }
    This scenario could represent a situation similar to the
    \qm{unrelenting influencers} scenario above, with two differences.
    First, one TV network has much higher influence than the other.
    Second, the networks are slightly influenced by all the agents
    (e.g., by checking ratings and choosing what news to cover accordingly).
\end{itemize}

\noindent
\review{We simulated the evolution of agents' beliefs
and the corresponding polarization of the network
for all combinations of initial belief configurations
and influence graphs presented above, using both the
confirmation-bias update-function (Definition~\ref{def:confirmation-bias})
and the classical update-function (Remark~\ref{rem:auto-bias}).
Simulations of circular influences used $n = 12$ agents,
whereas the rest used \review{$n = 1,000$} agents.
Both the Python implementation of the model and the Jupyter Notebook
containing the simulations are available on \review{GitHub}%
~\cite{website:github-repo}.
}

\review{The cases in which agents employ the confirmation-bias
update-function, which is the core of the present work,
are summarized in Figure~\ref{fig:comparing-num-bins}.
In that figure, each column corresponds to a different initial
belief configuration, and each row corresponds
to a different influence graph,
so we can visualize how the behavior
of polarization under confirmation bias changes when we fix an
influence graph (row) and vary the
initial belief configuration (column), or vice-versa.
In Section~\ref{sec:general-result} and Section~\ref{sec:specific-cases} we shall discuss some of these results in more detail when illustrating our formal results.
But first, in the following section we highlight some insights on the evolution
of polarization we obtain from the whole set of simulations.}

\review{
\begin{figure}[tbp]
\centering
\includegraphics[width=1\linewidth]{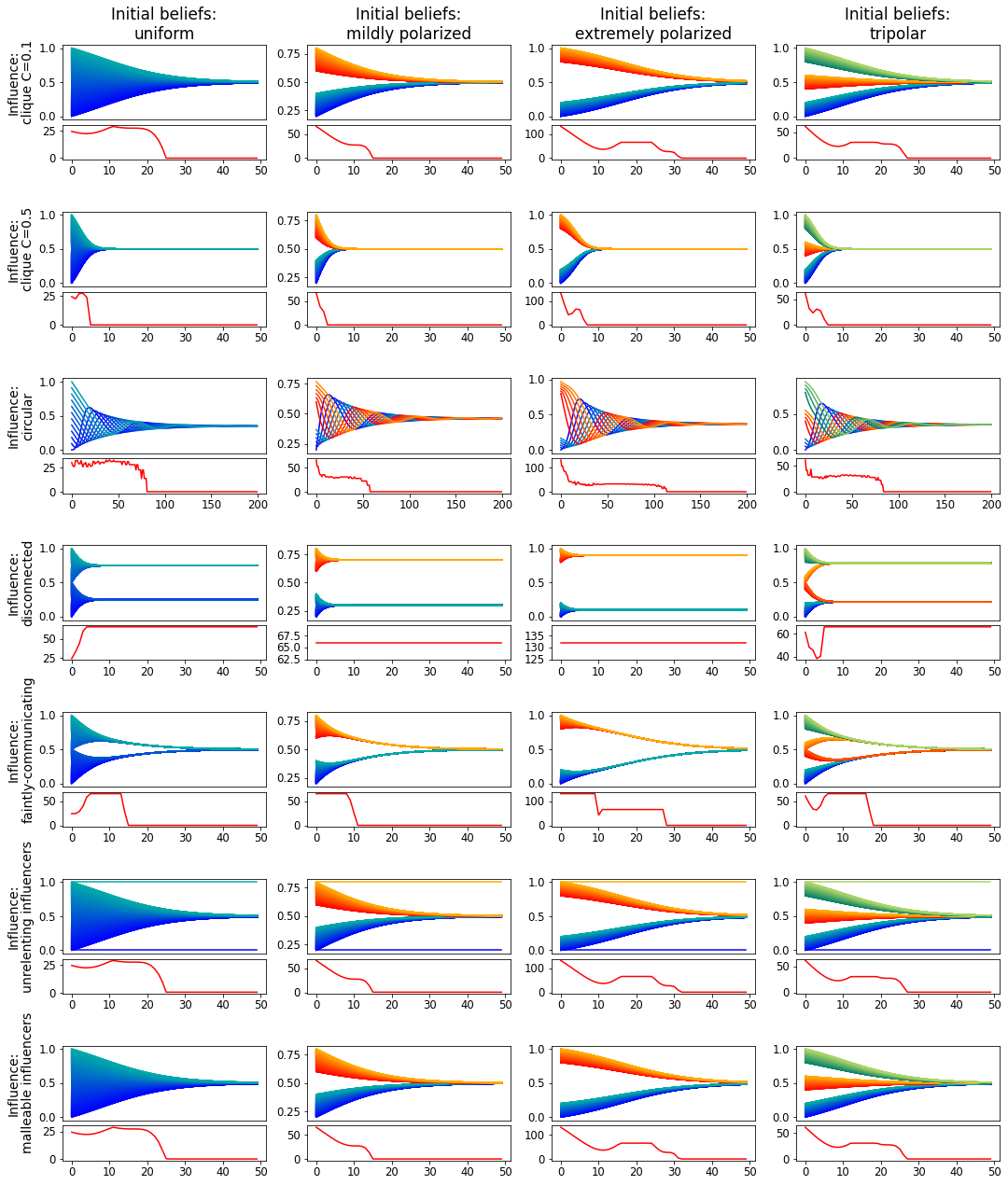}
  \caption{\review{
  Evolution of belief  and  polarization
  under
  confirmation-bias.
  Each row (resp.\ column) contains all graphs with the same influence graph (resp.\ initial belief configuration).
   Circular influences used $n=12$ agents, the rest used \review{$n=1,000$} agents.
  Each graph should be read as in Figure~\ref{running-example:figure}.
  Agents with similar behavior are grouped by colors:
  e.g., in all graphs in the rightmost column (tripolar initial belief configuration) all agents initiating in one of the three poles have
  the same color (green, orange, or blue).
}
  }%
  \label{fig:comparing-num-bins}
\end{figure}
}

\subsection{Insights from simulations}%
\label{sec:simulations-insights}
We divide our discussion into \qm{expected} and \qm{unexpected} behaviors identified.

\subsubsection{Analysis of \qm{expected} behavior of polarization.}
\begin{figure}[tbp]
            \centering
            \begin{subfigure}[t]{.48\linewidth}
                \centering
                \includegraphics[width=\linewidth]{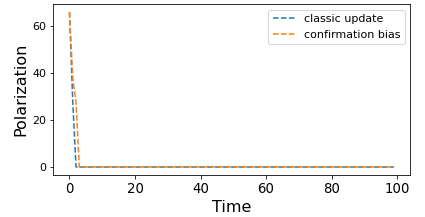}
                \caption{Comparison of update functions under the mildly polarized initial belief configuration, and the $C$-clique influence graph.}%
                \label{fig:simulations-expected-1}
            \end{subfigure}
            \hfill
            \begin{subfigure}[t]{.48\linewidth}
                \centering
                \includegraphics[width=\linewidth]{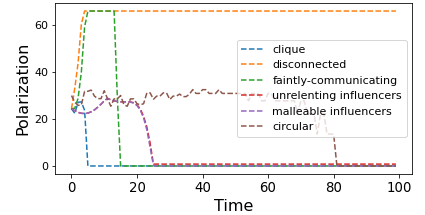}
                \caption{Comparison of influence graphs under the confirmation-bias update function, and the uniform initial belief configuration.}%
                \label{fig:simulations-expected-2}
            \end{subfigure}
          \\
          \vspace{4mm}
            \begin{subfigure}[t]{.50\linewidth}
                \centering
                \includegraphics[width=\linewidth]{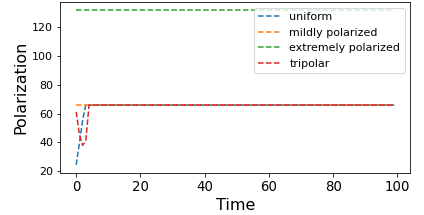}
                \caption{Comparison of initial belief \review{configurations} under the $C$-clique influence graph, and the classical update.
                }%
                \label{fig:simulations-expected-3}
            \end{subfigure}
            \caption{Examples of \qm{expected} \review{behavior of the evolution of polarization}.
            }%
            \label{fig:simulations-expected}
        \end{figure}

We start by considering the cases in which
our simulations agree with some perhaps \qm{expected} behavior
of polarization.
For this task, we focus on a scenario in which agents start off
with varied opinions, represented by the uniform initial belief configuration,
and all interact with each other via a $C$-clique influence graph.
We consider both the cases in which agents incorporate new
information in a classic way without confirmation bias (Remark~\ref{rem:auto-bias}), and in which agents present confirmation bias (Definition~\ref{def:confirmation-bias}).
These \qm{expected} results are shown in Figure~\ref{fig:simulations-expected}.

In particular, Figure~\ref{fig:simulations-expected-1} meets
our expectation that social networks in which all agents can
interact in a direct way eventually converge to a consensus
(i.e., polarization disappears),
even if agents start off with very different opinions and
are prone to confirmation bias.
Figure~\ref{fig:simulations-expected-2} shows that for the same
fixed update function and initial belief \review{configuration},
different interaction graphs may lead to very different evolutions
of polarization:
it grows to a maximum if agents are disconnected,
achieves a very low yet non-zero value in the presence of 2 unrelenting influencers,
and disappears in all other cases
(i.e., when agents can influence each other, even if indirectly).
%
Finally, Figure~\ref{fig:simulations-expected-3} shows that when
there are agents that do not communicate with each other at all,
as in a disconnected influence graph,
then even rational agents updating beliefs according to the classic
update function may not reach consensus, and polarization
may stabilize at relatively high values.

We notice that some of the expected behaviors described in this section \emph{will} follow from the results of in Section~\ref{sec:specific-cases} (see, e.g., Theorem~\ref{theorem:cb-geberal-scc-convergence}).

\subsubsection{Analysis of perhaps \qm{unexpected} behavior of polarization.}
We now turn our attention interesting cases in which the simulations help shed
light on perhaps counter-intuitive behavior of the dynamics of polarization.
In the following we organize these insights into several topics.

\subsubsection*{Polarization is not necessarily a monotonic function of time.}
\begin{figure}[tbp]
            \centering
            \begin{subfigure}[t]{.48\linewidth}
                \centering
                \includegraphics[width=\linewidth]{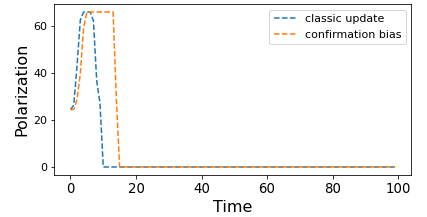}
                \caption{Comparison of update functions under the uniform initial belief configuration, and the faintly communicating influence graph.}%
                \label{fig:pol-non-monotonic-1}
            \end{subfigure}
            \hfill
            \begin{subfigure}[t]{.48\linewidth}
                \centering
                \includegraphics[width=\linewidth]{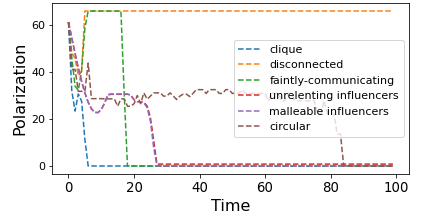}
                \caption{Comparison of interaction graphs under the confirmation-bias update, and the midly polarized initial belief configuration.
                }%
                \label{fig:pol-non-monotonic-2}
            \end{subfigure}
            \caption{Examples of cases in which the evolution of polarization is not monotonic.
             }%
            \label{fig:pol-non-monotonic}
        \end{figure}

At first glance it may seem that for a fixed social network
configuration, polarization either only increases or decreases
with time.
Perhaps surprisingly, this is not so, as illustrated in
Figure~\ref{fig:pol-non-monotonic}.

We start by noticing that Figure~\ref{fig:pol-non-monotonic-1} shows that
for a uniform initial belief \review{configuration} and an interaction graph
with two faintly communicating groups, no matter the update function,
polarization increases before decreasing again and stabilizing
at a lower level.
This is explained by the fact that in such a set-up agents
within the same group will reach consensus relatively fast,
which will lead society to two clear poles and increase polarization
at an initial stage.
However, as the two groups continue to communicate over time,
even if faintly, a general consensus is eventually achieved.

Figure~\ref{fig:pol-non-monotonic-2} shows that a tripolar
social network in which agents have confirmation bias,
a similar phenomenon occurs for all interaction graphs.
The only exceptions are the cases of disconnected groups, in which
polarization stabilizes at a high level, and of unrelenting influencers,
in which polarization stabilizes at a very low yet non-zero level.
In the first case, this happens because the disconnected groups reach internal consensus but remain far from the other group, since they do not communicate. This represents a high level of polarization. The case of two unrelenting influencers retains a low level of polarization simply because the opinions of the unrelenting influencers never change, even if the rest of agents attain \review{consensus}.
Another interesting case is that of two faintly communicating groups: here polarization first increases as the groups reach internal consensus, but then the two groups move toward one another and polarization decreases. The other two interaction graphs also stabilize to zero polarization.

\subsubsection*{The effect of different update functions.} 
\begin{figure}[tbp]
            \centering
            \begin{subfigure}[t]{.48\linewidth}
                \centering
                \includegraphics[width=\linewidth]{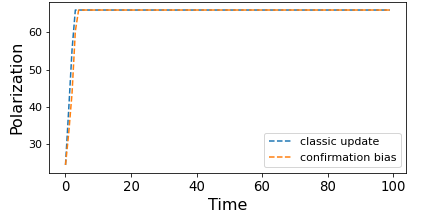}
                \caption{Comparison of update functions under the uniform initial belief configuration, and the disconnected influence graph.}%
                \label{fig:simulation-update-functions-1}
            \end{subfigure}
            \hfill
            \begin{subfigure}[t]{.48\linewidth}
                \centering
                \includegraphics[width=\linewidth]{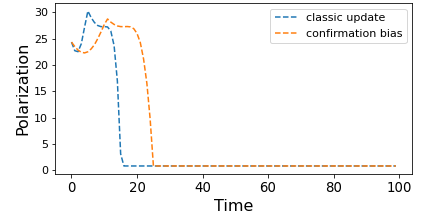}
                \caption{Comparison of update functions under the uniform belief configuration and the unrelenting-influencers influence graph.}%
                \label{fig:simulation-update-functions-2}
            \end{subfigure}
            \\
            \begin{subfigure}[t]{.48\linewidth}
                \centering
                \includegraphics[width=\linewidth]{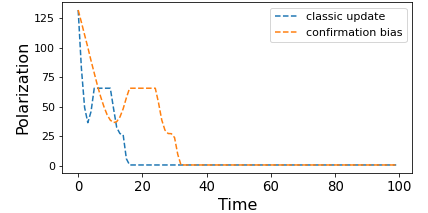}
                \caption{Comparison of update functions under the extremely-polarized initial belief configuration, and the unrelenting-influencers influence graph.}%
                \label{fig:simulation-update-functions-3}
            \end{subfigure}
            \caption{Comparison of different update functions.
            }%
            \label{fig:simulation-update-functions}
        \end{figure}

Figure~\ref{fig:simulation-update-functions} shows a comparison
of different update functions in various interesting scenarios.

In particular, Figure~\ref{fig:simulation-update-functions-1} shows that,
as expected, polarization can permanently increase in a disconnected
social network, with little difference between the behavior of different update functions.
Figure~\ref{fig:simulation-update-functions-2} depicts the effects of the two different update functions beginning from a uniform belief \review{configuration}, with two unrelenting influencers as the influence function. In both cases, all agents except the influencers eventually reach a belief value of 0.5 (the middle of the belief spectrum, between the two extreme agents), representing an increased but still fairly low level of polarization. The classic belief update function achieves this equilibrium fastest, since in under confirmation bias agents are less influenced by others whose beliefs are far from their own, so their beliefs change more slowly.
Finally, Figure~\ref{fig:simulation-update-functions-3} shows that even under two extreme unrelenting influencers,  consensus is eventually nearly reached, since everyone except the influencers eventually reaches a belief \review{configuration} between the beliefs of the two influencers.

\subsubsection*{The effect of different interaction graphs.}
\begin{figure}[tbp]
            \centering
            \begin{subfigure}[t]{.48\linewidth}
                \centering
                \includegraphics[width=\linewidth]{figures/CONFBIAS_X_UNIFORM-no-title.png}
                \caption{Comparison of interaction graphs under the confirmation-bias update, and the uniform initial belief configuration.}%
                \label{fig:simulation-interaction-graphs-3}
            \end{subfigure}
            \hfill
            \begin{subfigure}[t]{.48\linewidth}
                \centering
                \includegraphics[width=\linewidth]{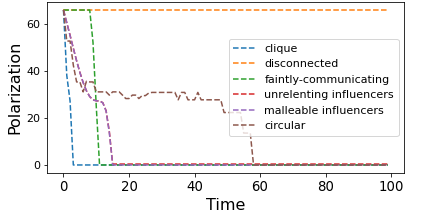}
                \caption{Comparison of interaction graphs under the confirmation-bias update, and the mildly polarized initial belief configuration.}%
                \label{fig:simulation-interaction-graphs-4}
            \end{subfigure}
            \caption{Comparison of different interaction graphs.
             }%
            \label{fig:simulation-interaction-graphs}
        \end{figure}

Figure~\ref{fig:simulation-interaction-graphs} shows a
comparison of different interaction graphs in various scenarios.
Figures~\ref{fig:simulation-interaction-graphs-3} and~\ref{fig:simulation-interaction-graphs-4} show that a faintly
communicating graph leads to a temporary peak in polarization,
which is reversed in all cases.
As we discussed, this is explained by the fact that agents within the same
group achieve consensus faster than agents in different groups,
leading to a temporary increase in polarization.
Note as well that both figures show that the presence of two unrelenting influencers pushing their agendas is sufficient prevent \review{consensus-reaching}, even if polarization remains at a very low level.

\subsubsection*{The effect of different initial belief \review{configurations}.} 
\begin{figure}[tbp]
            \centering
            \begin{subfigure}[t]{.48\linewidth}
                \centering
                \includegraphics[width=\linewidth]{figures/CLASSIC_GROUP_2_DISCONECTED_X-no-title.png}
                \caption{Comparison of initial belief \review{configurations} under the disconnected influence graph, and the classical update.}%
                \label{fig:simluation-belief-states-1}
            \end{subfigure}
            \hfill
            \begin{subfigure}[t]{.48\linewidth}
                \centering
                \includegraphics[width=\linewidth]{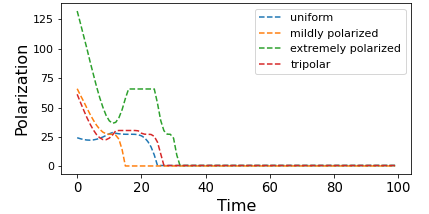}
                \caption{Comparison of initial belief \review{configurations} under the unrelentin-influencers influence graph, and the confirmation-bias update.}%
                \label{fig:simluation-belief-states-2}
            \end{subfigure}
            \caption{Comparison of different initial belief \review{configurations}.
            }%
            \label{fig:simluation-belief-states}
\end{figure}

Figure~\ref{fig:simluation-belief-states} compares different
belief \review{configurations} in various scenarios.
Figure~\ref{fig:simluation-belief-states-1} and
Figure~\ref{fig:simluation-belief-states-2} show that
even initial configurations with very different levels
of polarization can converge to a same polarization level
under both classic update and confirmation bias. However,
not all initial belief configurations converge to a same
final value, as is the case with the extremely polarized
curve in Figure~\ref{fig:simluation-belief-states-1}.


\section{Belief and Polarization Convergence}%
\label{sec:general-result}
Polarization tends to diminish as agents approximate a \emph{consensus}, i.e.,
as they (asymptotically) agree upon a common belief value for the proposition of interest.  Here and in Section~\ref{sec:specific-cases} we consider meaningful families of influence graphs that guarantee consensus \emph{under confirmation bias}.
We also identify fundamental properties of agents,
and the value of convergence.
Importantly, we relate influence with the notion of \emph{flow} in flow networks, and use it to identify necessary conditions for polarization not converging to zero.

\subsection{Polarization at the limit}

Proposition~\ref{pol-consensus} states that our polarization measure on
a belief configuration  (Definition~\ref{k-bin:def}) is zero exactly when all
belief values in it lie within the same bin of the underlying discretization
$D_k = I_0\ldots I_{k-1}$ of $[0,1]$.
In our model polarization converges to zero if all agents' beliefs converge
to a same non-borderline value. More precisely:

\begin{restatable}[Zero Limit Polarization]{lem}{respolatlimit}%
\label{pol-at-limit}
 Let $v$ be a non-borderline point of $D_k$ such that for every $\agent{i} \in  \Agents$, $\lim_{t \to \infty} \Bfun{i}{t} = v$. Then $\lim_{t \to \infty} \Pfun{\Blft{t}} = 0$.
\end{restatable}

\begin{figure}[tb]
    \centering
    \begin{subfigure}[t]{.40\textwidth}
     \centering
     \includegraphics[width=\linewidth]{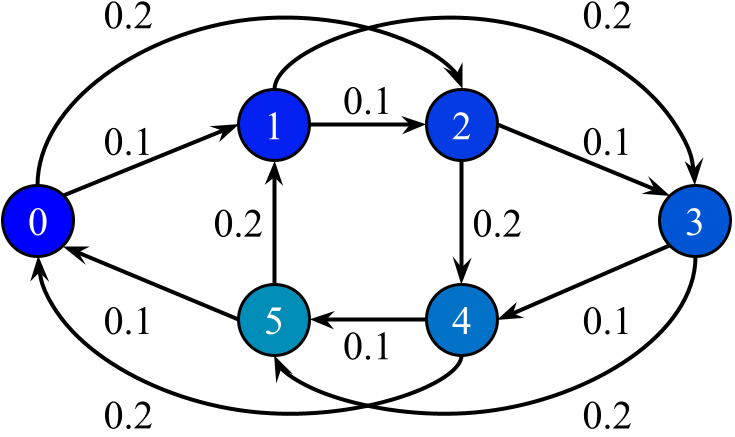}
     \caption{Influence graph.}%
     \label{fig:double-circular-graph}
    \end{subfigure}
    \\
    \vspace{2mm}
    \begin{subfigure}[t]{0.48\textwidth}
     \centering
     \includegraphics[width=\linewidth]{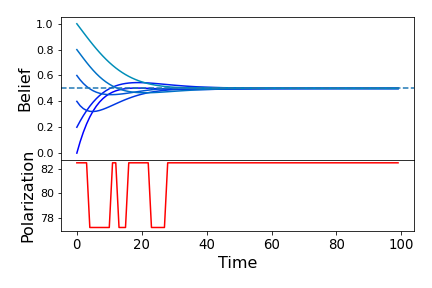}
     \caption{\review{Evolution of beliefs and polarization, 2 bins.}
     }\label{fig:borderline-2bin-polarization}
    \end{subfigure}
    \hfill
    \begin{subfigure}[t]{0.48\textwidth}
     \centering
     \includegraphics[width=\linewidth]{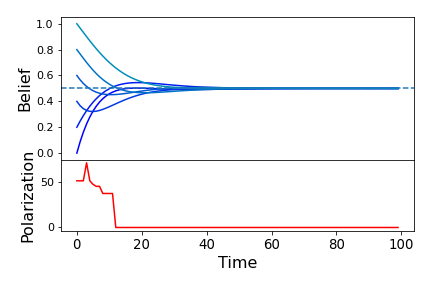}
     \caption{\review{Evolution of beliefs and polarization, 3 bins.}
     }\label{fig:borderline-3bin-polarization}
    \end{subfigure}
    \caption{Belief convergence to borderline value \review{$0.5$}. Polarization does not converge to 0 with equal-length 2 bins (Figure~\ref{fig:borderline-2bin-polarization}) but it does with 3 equal-length bins  (Figure~\ref{fig:borderline-3bin-polarization}).
    In all cases the initial belief values for Agents $\agent{0}$ to $\agent{5}$ are, respectively, $0.0$, $0.2$, $0.4$, $0.6$, $0.8$, and $1.0$ and the \review{influence graph is the one depicted in Figure~\ref{fig:double-circular-graph}}.
    \review{The graphs for the evolution of belief and polarization should be read as in Figure~\ref{running-example:figure}.}
    }%
    \label{fig:borderline-example}
\end{figure}

To see why we exclude the $k - 1$
borderline values of $D_k$ in the above lemma, assume $v \in I_m$ where $m \in \{0,\ldots,k-1\}$ is a borderline value. Suppose that there are two agents $i$ and $j$ whose beliefs converge to $v$,
but with the belief of $i$ staying always within $I_m$ whereas the belief of $j$ remains outside of $I_m$.
Under these conditions one can verify, using Definition~\ref{def:poler} and Definition~\ref{k-bin:def}, that $\Pol$ will not converge to $0$. This situation is illustrated in Figure~\ref{fig:borderline-2bin-polarization} assuming a discretization $D_2 = \ropen{0, \nicefrac{1}{2}}, [\nicefrac{1}{2},1]$ whose only borderline is $\nicefrac{1}{2}$. Agents' beliefs converge to value $v = \nicefrac{1}{2}$, but polarization does not converge to 0. In contrast,  Figure~\ref{fig:borderline-3bin-polarization} illustrates Lemma~\ref{pol-at-limit} for  $D_3= \ropen{0,\nicefrac{1}{3}}, \ropen{\nicefrac{1}{3},\nicefrac{2}{3}}, [\nicefrac{2}{3},1]$.
~\footnote{\ It is worthwhile to note that this discontinuity at borderline points matches real scenarios where each bin represents a sharp action an agent takes based on his current belief value.
Even when two agents' beliefs are asymptotically converging to a same borderline value from different sides, their discrete decisions will remain distinct.
E.g., in the vaccine case of Example~\ref{running-example}, even agents that are
asymptotically converging to a common belief value of $0.5$ will take different
decisions on whether or not to vaccinate, depending on which side of $0.5$ their belief falls.
In this sense, although there is convergence in the underlying belief values, there remains polarization \review{with respect to} real-world actions taken by agents.}

\subsection{Convergence under Confirmation Bias in Strongly Connected Influence}%
\label{sec:main-result}
We now introduce the family of \emph{strongly-connected} influence graphs, which includes cliques, that describes scenarios where each agent has an influence over all others. Such influence is not necessarily \emph{direct} in the sense defined next, or the same for all agents, as in the more specific cases of cliques.

\begin{defi}[Influence Paths]\label{def:influence-path} Let $C \in \lopen{0,1}$.
We say that $\agent{i}$ has a \emph{direct influence} $C$ over $\agent{j}$, written  $\ldinfl{\agent{i}}{C}{\agent{j}}$, if $\Ifun{\agent{i}}{\agent{j}}=C$.

 An \emph{influence path} is a {finite sequence} of \emph{distinct} agents from $\Agents$ where each agent in the sequence has a direct influence over the next one. Let  $p$ be an influence path $i_0i_1\ldots i_n$. The \emph{size} of $p$ is $|p| = n$.
We also use $\ldinfl{\agent{i_0}}{C_1}{\agent{i_1}}\ldinfl{}{C_2}{}\ldots \ldinfl{}{C_n} {\agent{i_n}}$ to denote  $p$ with the direct influences along this path. We write  $\linfl{\agent{i_0}}{C}{p}{\agent{i_n}}$ to indicate that the \emph{product influence} of $i_0$ over $i_n$ along  $p$ is $C = C_1 \times \cdots  \times C_n$.

 We often omit influence or path indices from the above arrow notations when they are unimportant or clear from the context.  We say that   $\agent{i}$ has an \emph{influence} over $\agent{j}$ if $\infl{\agent{i}}{\agent{j}}$.
\end{defi}

The next definition is akin to the graph-theoretical notion of strong connectivity.
\begin{defi}[Strongly Connected Influence]\label{def:strongly-connected}  We say that an influence graph $\Inter$ is \emph{strongly connected} if for all $\agent{i}$, $\agent{j} \in  \Agents$ such that $i{\neq}j$, $\infl{\agent{i}}{\agent{j}}$.
\end{defi}

\begin{rem}\label{rmk:assumption}  For technical reasons we assume that, \emph{initially}, there are no two agents $\agent{i}, \agent{j} \in \Agents$ such that $\Bfun{i}{0} =  0$ and $\Bfun{j}{0} = 1$. This implies that
 for every $\agent{i}, \agent{j} \in \Agents$: $\CBfun{\agent{i}}{\agent{j}}{0} >  0$ where  $\CBfun{\agent{i}}{\agent{j}}{0}$ is the confirmation bias of $i$ towards $j$ at time $0$ (See Definition~\ref{def:confirmation-bias}). Nevertheless, at the end of this section we will address the cases in which this condition does not hold.
\end{rem}

We shall use the following notation for the extreme beliefs of agents.

\begin{defi}[Extreme Beliefs]\label{def:extreme:beliefs} Define   $\mx{t}=\max_{\agent{i} \in \Agents} \Bfun{\agent{i}}{t}$ and $\mn{t}= \max_{\agent{i} \in \Agents} \Bfun{\agent{i}}{t}$ as the functions giving the maximum and minimum belief values, respectively, at time $t$.
 \end{defi}

It is worth noticing that \emph{extreme agents} --i.e., those holding extreme beliefs-- do not necessarily remain the same across time steps.
\review{Figure~\ref{fig:circulation-graph-a}}
illustrates this point:
Agent 0 goes from being the one most against the proposition of interest
at time $t = 0$ to being the one most in favour of it around $t = 8$.
Also, the third row of Figure~\ref{fig:comparing-num-bins} shows simulations
for a circular graph under several initial belief configurations.
Note that under all initial belief configurations
different agents alternate as maximal and minimal belief holders.

Nevertheless, in what follows will show that the beliefs of all agents, under strongly-connected influence  and confirmation bias, converge to the same value since the difference between $\mn{t}$ and $\mx{t}$ goes to 0 as $t$ approaches infinity.
We begin with a  lemma stating a property of the confirmation-bias update: \emph{The belief value of any agent at any time is bounded by those from extreme agents in the previous time unit}.

\begin{restatable}[Belief Extremal Bounds]{lem}{reslemmacbmaxdiffmin}%
\label{lemma:cb-maxdiffmin}
 For every $i\in \Agents$,   $\mn{t} \leq \Bfun{\agent{i}}{t + 1} \leq \mx{t}$.
\end{restatable}

The next corollary follows from the assumption in Remark~\ref{rmk:assumption} and Lemma~\ref{lemma:cb-maxdiffmin}.

\begin{restatable}[]{cor}{rescorbiasfactor}%
\label{cor:biasfactor}
For every $\agent{i}, \agent{j} \in \Agents$, $t \geq 0$: $\CBfun{\agent{i}}{\agent{j}}{t} > 0$.
\end{restatable}

Note that monotonicity does not necessarily hold for belief evolution. This
is illustrated by Agent 0's behavior in
\review{Figure~\ref{fig:circulation-graph-a}.}
However, it follows immediately from Lemma~\ref{lemma:cb-maxdiffmin} that $\mn{}$ and $\mx{}$ in Definition~\ref{def:extreme:beliefs} are monotonically non-decreasing and non-increasing functions of $t$.

\begin{restatable}[Monotonicity of Extreme Beliefs]{cor}{rescorcbmbeforemafter}\label{cor:cb-mbefore-mafter}
$\mx{t+1} \leq \mx{t}$ and $\mn{t+1} \geq \mn{t}$ for all $t \in \nat$.
\end{restatable}

Monotonicity and the bounding of $\mx{\cdot}$, $\mn{\cdot}$ within $[0,1]$ lead us, via the Monotonic Convergence Theorem~\cite{Sohrab:14}, to the existence of \emph{limits for beliefs of extreme agents}.

\begin{restatable}[Limits of Extreme Beliefs]{thm}{rescorcbmaxlimitsexist}%
\label{th:cb-max-limits-exist} There are $U,L \in [0,1]$ \review{such that}
$\lim_{t\to\infty} \mx{t} = U$ and $\lim_{t\to\infty} \mn{t} = L$.
\end{restatable}

We still need to show that $U$ and $L$ are the same value. For this we prove a distinctive property of agents under strongly connected  influence graphs: the belief of any agent at time $t$ will influence every other agent by the time $t + |\Agents| - 1$. This is precisely formalized below in Lemma~\ref{lemma:cb-path-bound}.
First, however, we introduce some bounds for confirmation-bias and influence, as well as notation for the limits in Theorem~\ref{th:cb-max-limits-exist}.

\begin{defi}[Min Factors]\label{def:fcb-min}
Define $\CBfunM = \min_{\agent{i}, \agent{j} \in \Agents} \CBfun{i}{j}{0}$ as the minimal confirmation bias factor at $t = 0$. Also let $\IfunM$ be the smallest positive influence in $\Inter$.  Furthermore, let
$L = \lim_{t\to\infty} \mn{t}$ and $U = \lim_{t\to\infty} \mx{t}$.
\end{defi}

Notice that since $\mn{t}$ and $\mx{t}$ do not get further apart as the time $t$ increases (Corollary~\ref{cor:cb-mbefore-mafter}),  $\min_{\agent{i}, \agent{j} \in  \Agents} \CBfun{i}{j}{t}$  is a non-decreasing function of $t$. Therefore $\CBfunM$  acts as a lower bound for the confirmation-bias factor in every time step.
\begin{restatable}[]{prop}{resfcbminprop}%
\label{fcb-min:prop}
$\CBfunM =  \min_{\agent{i}, \agent{j} \in \Agents} \CBfun{i}{j}{t} $ for every $t > 0$.
\end{restatable}

We use the factor $\CBfunM$ in the next result to establish that the belief of agent $\agent{i}$ at time $t$, the minimum confirmation-bias factor, and the maximum belief at $t$ act as bound of the belief of $\agent{j}$ at $t + |p|$,
where $p$ is an influence path from $\agent{i}$ and $\agent{j}$.

\begin{restatable}[Path bound]{lem}{reslemmacbpathbound}%
\label{lemma:cb-path-bound}
If $\Inter$ is strongly connected:
\begin{enumerate}
    \item Let  $p$ be an arbitrary path $\linfl{\agent{i}}{C}{p}{\agent{j}}$.  Then
    \[\Bfun{\agent{j}}{t+|p|} \leq \mx{t} + \nicefrac{C\CBfunM^{|p|}}{|\Agents|^{|p|}}(
\Bfun{\agent{i}}{t} - \mx{t}).\]

    \item  Let $\mstar^t \in \Agents$ be an agent holding the least belief value at time $t$ and $p$ be a path such that $\linfl{\agent{\mstar}^t}{}{p}{\agent{i}}$.
    Then
    \[\Bfun{\agent{i}}{t + |p|} \leq \mx{t} - \delta,\]
    with $\delta = \left(\nicefrac{\IfunM\CBfunM}{|\Agents|}\right)^{|p|}(U - L)$.
\end{enumerate}
\end{restatable}

\noindent
Next we establish that all beliefs at time $t + |\Agents| - 1$ are smaller
than the maximal belief at $t$ by a factor of at least $\epsilon$ depending
on the minimal confirmation bias, minimal influence and the limit values $L$
and $U$.

\begin{restatable}[]{lem}{reslemcbepsilonbound}%
\label{lem:cb-epsilon-bound}
Suppose that $\Inter$ is strongly-connected.
\begin{enumerate}
    \item If $\Bfun{\agent{i}}{t + n} \leq \mx{t} - \gamma$ and $\gamma \geq 0$ then $\Bfun{\agent{i}}{t+n+1} \leq \mx{t} - \nicefrac{\gamma}{|\Agents|}$.

    \item $\Bfun{\agent{i}}{t+|\Agents|-1} \leq \mx{t} - \epsilon$, where $\epsilon$ is equal to $\left(\nicefrac{\IfunM\CBfunM}{|\Agents|}\right)^{|\Agents|-1}(U-L)$.
\end{enumerate}
\end{restatable}

\noindent
Lemma~\ref{lem:cb-epsilon-bound}(2) 
states that $\max^{\cdot}$ decreases by at least $\epsilon$ after  $|A| - 1$ steps. Therefore, after $m(|A|-1)$ steps it should decrease by at least $m\epsilon$.

\begin{restatable}[]{cor}{rescormaxdiff}%
\label{cor:max-dif}
If $\Inter$ is strongly connected,
$\mx{t+m(|\Agents|-1)} \leq \mx{t} - m\epsilon$ for $\epsilon$ in  Lemma~\ref{lem:cb-epsilon-bound}.
\end{restatable}

We can now state that in strongly connected influence graphs extreme beliefs eventually converge to the same value. The proof uses Corollary~\ref{cor:biasfactor} and Corollary~\ref{cor:max-dif} above.

\begin{restatable}[]{thm}{resthul}%
\label{th:U=L}
If $\Inter$ is strongly connected  then
$\lim_{t\to\infty} \mx{t} =  \lim_{t\to\infty} \mn{t}$.
\end{restatable}

Combining Theorem~\ref{th:U=L}, the assumption in  Remark~\ref{rmk:assumption} and the Squeeze Theorem~\cite{Sohrab:14}, we conclude that for strongly-connected graphs, all agents' beliefs converge to the same value.

\begin{restatable}[]{cor}{restheoremcbsccconvergence}%
\label{cor:cb-scc-convergence}
If $\Inter$ is strongly connected then for all
$\agent{i},\agent{j} \in \Agents, \lim_{t \to \infty} \Bfun{\agent{i}}{t}  = \lim_{t \to \infty} \Bfun{\agent{j}}{t}$.
\end{restatable}

\subsubsection{The Extreme Cases.}\label{extreme:case} We assumed in Remark~\ref{rmk:assumption} that there were no two agents $\agent{i},\agent{j}$ \review{such that} $\Bfun{\agent{i}}{t} = 0$ and $\Bfun{j}{t} = 1$.
Theorem~\ref{theorem:cb-geberal-scc-convergence} below addresses
the situation in which this does not happen.
More precisely, it establishes that under confirmation-bias update,
in any strongly-connected, non-radical society, agents'
beliefs eventually converge to the same value.

\begin{defi}[Radical Beliefs]\label{radical:def}
An agent $\agent{i} \in \Agents$ is called \emph{radical} if $\Blf_{\agent{i}} = 0$ or $\Blf_{\agent{i}} = 1$.
A belief configuration $\Blf$ is \emph{radical} if every $\agent{i} \in \Agents$ is radical.
\end{defi}

\begin{restatable}[Confirmation-Bias Belief Convergence]{thm}{restheoremcbgeberalsccconvergence}%
\label{theorem:cb-geberal-scc-convergence}
In a strongly connected influence graph and under the confirmation-bias update-function, if $\Blft{0}$ is not radical then for all $\agent{i}, \agent{j}  \in \Agents$, $\lim_{t \to \infty} \Bfun{i}{t} = \lim_{t \to \infty} \Bfun{j}{t}$. Otherwise for every $\agent{i} \in \Agents$, $\Bfun{i}{t} = \Bfun{i}{t+1} \in \{0,1\}$.
\end{restatable}

We conclude this section by emphasizing that belief convergence is not guaranteed
in non strongly-connected graphs.
Figure~\ref{fig:weakly-graph} from the vaccine example shows such
a graph where neither belief convergence nor zero-polarization is
obtained.

\section{Conditions for Polarization}%
\label{sec:specific-cases}
We now use concepts from flow networks to  identify insightful
necessary conditions for polarization never disappearing.  Understanding the conditions when polarization
\emph{does not} disappear under confirmation bias is one of the main contributions of this paper.

\subsection{Balanced Influence: Circulations}

The following notion is inspired by the \emph{circulation problem} for directed graphs (or flow network)~\cite{Diestel:17}.
Given a graph $G=(V,E)$ and a function $c : E \to  \reals$ (called \emph{capacity}), the problem involves finding a function  $f : E \to  \reals$ (called \emph{flow}) such that:
\begin{enumerate}
    \item $f(e) \leq c(e)$ for each $e \in  E$; and

    \item $\sum_{(v,w) \in  E}f(v,w)=\sum_{(w,v) \in E}f(w,v)$ for all $v \in V$.
\end{enumerate}
Such an $f$ exists is called a \emph{circulation} for $G$ and $c$.

Thinking of flow as influence, the second condition, called \emph{flow conservation}, corresponds to requiring that each agent influences others as much as is influenced by them.

\begin{defi}[Balanced Influence]\label{def:circulation} We say that  $\Inter$ is \emph{balanced} (or a \emph{circulation}) if every $\agent{i} \in \Agents$ satisfies the constraint $\sum_{\agent{j} \in \Agents} \Ifun{i}{j} =   \sum_{\agent{j} \in \Agents} \Ifun{j}{i}$.
\end{defi}

Cliques and circular graphs, where all (non-self) influence values are equal, are balanced (see Figure~\ref{fig:interaction-graphs-circular}).
The graph of our vaccine example (Figure~\ref{running-example:figure}) is a circulation that is neither a clique nor a circular graph.
Clearly, influence graph $\Inter$ is balanced if it is a solution to a circulation problem for some $G =   (\Agents,\Agents \times    \Agents)$ with capacity $c : \Agents \times    \Agents \to  [0,1]$.

Next we use a fundamental property from flow networks describing flow conservation for graph cuts~\cite{Diestel:17}. Interpreted in our case it says that any group of agents $A{\subseteq}\Agents$ influences other groups as much as they influence $A$.

\begin{restatable}[Group Influence Conservation]{prop}{respropgroupinfluenceconservation}%
\label{prop:group-influence-conservation}
Let $\Inter$ be balanced and $\{A,B\}$ be a partition of $\Agents$.
Then
\[\sum_{i \in A}\sum_{j \in B} \Ifun{i}{j} =\sum_{i \in A}\sum_{j \in B} \Ifun{j}{i}.\]
\end{restatable}

We now define \emph{weakly connected influence}.
Recall that an undirected graph is \emph{connected} if there is path between each pair of nodes.

\begin{defi}[Weakly Connected Influence]\label{def:weakly-connected}
Given an influence graph $\Inter$, define the undirected graph $G_{\Inter} =   (\Agents,E)$ where $\{i,j \} \in E$ if and only if $\Ifun{i}{j} >  0$ or $\Ifun{j}{i} >  0$.
An influence graph $\Inter$ is called \emph{weakly connected} if  the undirected graph $G_{\Inter}$ is connected.
\end{defi}

Weakly connected influence relaxes its strongly connected counterpart.
However, every balanced, weakly connected influence is strongly connected as implied by the next lemma. Intuitively, circulation flows never leaves strongly connected components. 

\begin{restatable}[]{lem}{resprocirculationpath}%
\label{prop:circulation-path}
If $\Inter$ is balanced and $\Ifun{i}{j} >  0$ then $\Path{j}{i}$.
\end{restatable}

\subsection{Conditions for Polarization}
We have now all elements to identify conditions for permanent polarization. The  convergence for strongly connected graphs (Theorem~\ref{theorem:cb-geberal-scc-convergence}), the polarization at the limit lemma (Lemma~\ref{pol-at-limit}), and Lemma~\ref{prop:circulation-path} yield the following noteworthy result.

\begin{restatable}[Conditions for Polarization]{thm}{respolnonzero}%
\label{pol-non-zero}
Suppose that $\lim_{t \to \infty}\Pfun{\Blft{t}}{\neq}0$. Then either:
\begin{enumerate}
    \item $\Inter$ is not balanced;
    \item $\Inter$ is not weakly connected;
    \item $\Blft{0}$ is radical; or
    \item for some borderline value $v$, $\lim_{t \to \infty} \Bfun{i}{t} =   v$ for each $\agent{i} \in \Agents$.
\end{enumerate}
\end{restatable}

\noindent
Hence, at least one of the four conditions is necessary for the persistence of polarization.
If (1) then there must be at least one agent that influences more than what he is influenced (or vice versa). This is illustrated in Figure~\ref{fig:weakly-graph} from the vaccine example, where Agent 2 is such an agent.
If (2) then there must be isolated subgroups of agents; e.g., two isolated strongly-connected components where the members of the same component will achieve consensus but the consensus values of the two components may be very different. This is illustrated in the fourth row of Figure~\ref{fig:comparing-num-bins}. Condition (3) can be ruled out if there is an agent that is not radical, like in all of our examples and simulations. As already discussed, (4) depends on the underlying discretization $D_k$ (e.g., assuming equal-length bins if $v$ is  borderline in $D_k$ it is not borderline in $D_{k+1}$, see Figure~\ref{fig:borderline-example}.).

\subsection{Reciprocal and Regular Circulations}%
\label{circulation:section}
The notion of circulation allowed us to identify potential causes of polarization. In this section we will also use it to identify meaningful topologies whose symmetry can help us predict the \emph{exact belief value} of convergence.

A \emph{reciprocal} influence graph is a 
circulation where the influence of $i$ over $j$ is the
same as that of $j$ over $i$, i.e, $\Ifun{i}{j} =    \Ifun{j}{i}$.  Also a graph is (\emph{in-degree})
\emph{regular} if the in-degree
of each nodes is the same;  i.e., for all $i,j \in \Agents$,
$|\Agents_{\agent{i}}| =   |\Agents_{\agent{j}}|$.

As examples of regular and reciprocal graphs, consider a graph $\Inter$  where all (non-self) influence values are equal. If $\Inter$ is \emph{circular} then it is a regular circulation, and if $\Inter$
is a \emph{clique} then it is a reciprocal regular circulation. Also we can modify slightly our vaccine example to obtain a regular reciprocal circulation as shown in Figure~\ref{fig:circulation-reg-rec}.


\review{
\begin{figure}[tb]
  \centering
   \raisebox{0.10\height}{\includegraphics[width=0.45\linewidth]{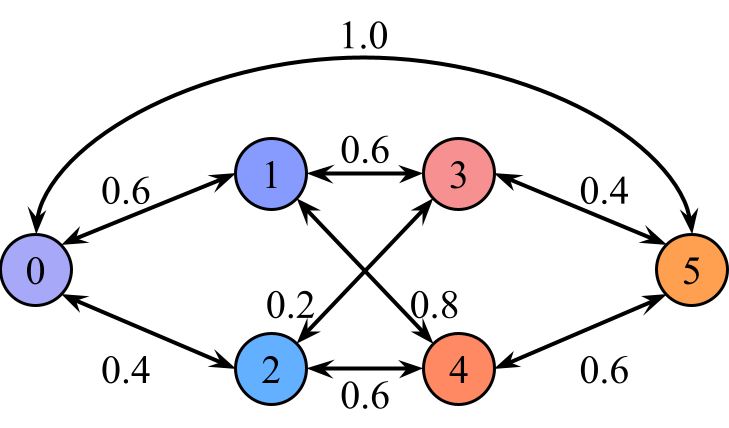}}
    \hfill
   \includegraphics[width=0.45\linewidth]{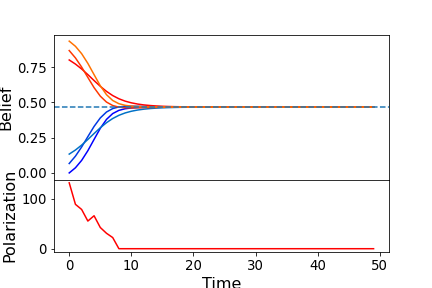}
   \caption{\review{Regular and reciprocal influence, and the corresponding evolution
   of beliefs and polarization. This figure should be read as Figure~\ref{running-example:figure}.}}%
   \label{fig:circulation-reg-rec}
\end{figure}
}

The importance of regularity and reciprocity of influence graphs is that their symmetry is sufficient to the determine the exact value all the agents converge to under
confirmation bias: \emph{the average of initial beliefs}.  Furthermore, under classical update (see Remark~\ref{rem:auto-bias}), we can drop reciprocity and obtain the same result. The result follows from Lemma~\ref{prop:circulation-path}, Theorem~\ref{theorem:cb-geberal-scc-convergence}, Corollary~\ref{cor:degroot}, the \review{Squeeze Theorem~\cite{Sohrab:14}} and by showing that
$\sum_{\agent{i} \in \Agents} \Bfun{i}{t} =   \sum_{\agent{i} \in  \Agents} \Bfun{i}{t+1}$ using symmetries derived from reciprocity, regularity, and the fact that $\CBfun{i}{j}{t} =   \CBfun{j}{i}{t}$.

\begin{restatable}[Consensus Value]{thm}{rescorcirculationconvergence}%
\label{cor:circulation-convergence}
Suppose that $\Inter$ is regular and weakly connected. If $\Inter$ is reciprocal and the belief update is confirmation-bias, or if the influence graph $\Inter$ is a circulation and the belief update is classical, then,
for every $i \in \Agents$,
\[\lim_{t \to \infty} \Bfun{i}{t} = \frac{1}{|\Agents|}\sum_{\agent{j} \in \Agents}\Bfun{j}{0}.\]
\end{restatable}

\section{Related Work}%
\label{sec:related-work}

\subsection{Comparison to DeGroot's model}%
\label{sec:degroot}
DeGroot proposed a very influential model, closely related to our work, to reason about learning and consensus  in multi-agent systems~\cite{degroot}, in which beliefs are updated by a constant stochastic matrix at each time step.
More specifically, consider
a group $\{1,2,\ldots,k\}$ of $k$ agents, \review{such that} each agent
$i$ holds an initial (real-valued) opinion $F_{i}^{0}$ on a given proposition
of interest.
Let $T_{i,j}$ be a non-negative weight that agent $i$ gives to agent $j$'s opinion,
\review{such that} $\sum_{j=1}^k T_{i,j}{=}1$.
DeGroot's model posits that an agent $i$'s opinion $F_{i}^{t}$ at any time
$t{\geq}1$ is updated as
$F_{i}^{t}{=}\sum_{j{=}1}^k T_{i,j} F_{i}^{t-1}$.
Letting $F^{t}$ be a vector containing all agents' opinions at time $t$,
the overall update can be computed as $F^{t{+}1}{=}T F^{t}$,
where $T{=}\{T_{i,j}\}$ is a stochastic matrix.
This means that the $t$-th configuration (for $t{\geq}1$) is
related to the initial one by $F^{t}{=}T^{t}F^{0}$, which is
a property thoroughly used to derive results in the model.

When we use classical update
(as in Remark~\ref{rem:auto-bias}), our model reduces to DeGroot's
via the transformation
$F_{i}^{0}{=}\Bfun{i}{0}$,
and
$T_{i,j}{=}\nicefrac{1}{|\Agents_i|} \ \Ifun{j}{i}$ if $i{\neq}j$,
or
$T_{i,j}{=}1{-}\nicefrac{1}{|\Agents_i|}\sum_{j{\in}\Agents_i} \Ifun{j}{i}$ otherwise.
Notice that $T_{i,j}{\leq}1$ for all $i$ and $j$, and, by construction, $\sum_{j{=}1}^k T_{i,j}{=}1$ for all $i$.
The following result is an immediate consequence of this reduction.
\begin{restatable}{cor}{rescordegroot}%
\label{cor:degroot}
    In a strongly connected influence graph $\Ifun{}{}$ and under the classical update function, for all $i, j{\in}\Agents$,
    \[\lim_{t{\to}\infty} \Bfun{i}{t} = \lim_{t{\to}\infty} \Bfun{j}{t}.\]
\end{restatable}

Unlike its classical counterpart, however, the confirmation-bias update
(Definition~\ref{def:confirmation-bias}) does not have an immediate correspondence with
DeGroot's model.
Indeed, this update is not linear due the confirmation-bias factor $\CBfun{i}{j}{t}{=}1{-}|\Bfun{j}{t}{-}\Bfun{i}{t}|$.
This means that in our model there is no immediate analogue of the
relation among arbitrary configurations and the initial one as the
relation in DeGroot's model (i.e., $F^{t}{=}T^{t}F^{0}$).
Therefore, proof techniques usually used in DeGroot's model
(e.g., based on Markov properties) are not immediately applicable
to our model.
In this sense our model is an extension of DeGroot's, and we need
to employ different proof techniques to obtain our results.

\review{The Degroot-like models are also used in~\cite{naive}. Rather than examining polarization and opinions, this work is concerned with the network topology conditions under which agents with noisy data about an objective fact converge to an accurate consensus, close to the true state of the world. As already discussed the basic DeGroot model does not include confirmation bias, however~\cite{sikder,mao,mf,hk, robust} all generalize DeGroot-like models to include functions that can be thought of as modelling confirmation bias in different ways, but with either no measure of polarization or a simpler measure than the one we use.~\cite{moreau} discusses DeGroot models where the influences change over time, and~\cite{survey} presents results about generalizations of these models, concerned more with consensus than with polarization. }

\subsection{Other related work}%
\label{sec:related-work-other}
We summarize some other relevant approaches and put into perspective the novelty of our approach.

\subsubsection*{Polarization} 
Polarization was originally studied as a psychological phenomenon in~\cite{M76},
and was first rigorously and quantitatively defined by economists Esteban and Ray~\cite{Esteban:94:Econometrica}. Their measure of polarization, discussed  in  Section~\ref{sec:model}, is influential, and we adopt it in this paper.
Li et al.~\cite{li}, and later Proskurnikov et al.~\cite{proskurnikov} modeled consensus and polarization in social networks. Like much other work, they treat polarization simply as the lack of consensus and focus on  when and under what conditions a population reaches \review{an agreement}\review{.
Elder's work}~\cite{alexis} focuses on methods to avoid polarization, without using a quantitative definition of polarization.

\subsubsection*{Formal Models} 
S{\^\i}rbu et al.~\cite{sirbu}
use a model  that updates probabilistically to investigate the effects of algorithmic bias on polarization by counting the number of opinion clusters, interpreting a single opinion cluster as consensus.
Leskovec et al.~\cite{gargiulo} simulate social networks and observe group formation over time. \review{Though belief update and group formation are related to our work~\cite{sirbu,gargiulo} are not concerned with a measure for polarization.}

\subsubsection*{Logic-based approaches}  
Liu et al.~\cite{liu} use ideas from doxastic and dynamic epistemic logics to
qualitatively model influence and belief change in social networks.
Seligman et al.~\cite{fblogic,facebook} introduce a basic ``Facebook logic.'' This logic is non-quantitative, but its interesting point is that an agent's possible worlds are different social networks. This is a promising approach to formal modeling of  epistemic issues in social networks.  Christoff~\cite{zoe} extends facebook logic and develops several non-quantitative logics for social networks, concerned with problems related to polarization, such as information cascades.
Young Pederson et al.~\cite{myp, myp2, myp3} develop a logic of polarization, in terms of positive and negeative links between agents, rather than in terms of their quantitative beliefs.
\review{Baltag et al.~\cite{bcrs} develop a dynamic epistemic logic of threshold models with imperfect information. In these models, agents either completely believe or completely disbelieve a proposition, and they update their belief over time based on the proportion of their neighbors that believe the proposition. Although this work is not concerned with polarization, it would be interesting to compare the situations where our model converges to the longterm outcomes of their threshold models, and consider the implications of the epistemic logic developed in their paper for our model. }
Hunter~\cite{hunter} introduces a logic of belief updates over social networks where closer agents in the social network are more trusted and thus more influential. While beliefs in this logic are non-quantitative, there is a quantitative notion of influence
between users.

\subsubsection*{Work on Users' Influence} 
The seminal paper Huberman et al.~\cite{huberman} is about determining which friends or followers in a user's network have the most influence on the user. Although this paper does not quantify influence between users, it does address an important question to our project. Similarly,~\cite{demarzo} focuses on finding most influential agents. This work on highly influential agents is relevant to our finding that such agents can maintain a network's polarization over time.  

\review{Work on decentralized gossip protocols  also examines the problem of information diffusion among agents from a different perspective~\cite{gossip1, gossip2, gossip3,gossip4}. The goal of a gossip protocol is for all the agents in a network to share all their information with as few communication steps as possible, using their own knowledge to choose their communication actions at each step.  It may be possible to generalize results from gossip protocols in order to understand how quickly it is possible for a social network to reach consensus under certain conditions. }

\section{Conclusions and Future Work}%
\label{sec:conclusion}

\review{In this paper we proposed a model for polarization and belief evolution for multi-agent systems under \emph{confirmation bias}; a recurrent and systematic pattern of deviation from rationality when forming opinions in social networks and in society in general. We extended previous results in the literature by showing that also under confirmation bias, whenever all agents can directly or indirectly influence each other, their beliefs always converge, and so does polarization as long as the convergence value is not a borderline point.  Nevertheless, we believe that our main contribution to the study of polarization is understanding how outcomes are structured when convergence does not obtain and polarization persists. Indeed, we have identified necessary conditions when polarization does not  disappear, and the convergence values for some important network topologies.}

\review{As future work we plan to explore the following directions aimed at making the present model more robust.}

\review{
\subsection{Dynamic Influence}
   In the current work, we consider one agent's influence on another to be static. For modelling short term belief change, this is a reasonable decision, but in order to model long term belief change in social networks over time, we need to include dynamic influence. In particular, agent $a$'s influence on agent $b$ should become stronger over time if agent $a$ sees $b$ as reliable, which means that $b$ mostly sends messages that are already close to the beliefs of agent $a$. Thus, we plan enrich our model with \emph{dynamic} influence between agents: agent $a$'s influence on $b$ becomes stronger if $a$ communicates messages that $b$ agrees with, and it becomes weaker if $b$ mostly disagrees with the beliefs that are communicated by $a$.  We expect that this change to the model will tend to increase group polarization, as agents who already tend to agree will have increasingly stronger influence on one another, and agents who disagree will influence one another less and less. This will be particularly useful for modelling \emph{filter bubbles}~\cite{filter}, and we will consult empirical work on this phenomenon in order to correctly include influence update in our model.
}
\review{
\subsection{Multi-dimensional beliefs} The current paper considers the simple case where agents only have beliefs about one proposition. This simplification may be quite useful for modelling some realistic scenarios, such as beliefs along a liberal/conservative political spectrum, but we believe that it will not be sufficient for modelling all situations. To that end, we plan to develop models where agents have beliefs about multiple issues.  This would allow us to represent, for example, Nolan's two-dimensional spectrum of political beliefs, where individuals can be economically liberal or conservative, and socially liberal or conservative~\cite{nolan}, since more nuanced belief situations such as this one cannot be modelled in a one-dimensional space. Since Esteban-Ray polarization considers individuals along a one-dimensional spectrum, we will extend Esteban-Ray polarization to more dimensions, or develop a new, multi-dimensional definition of polarization. }

\review{
\subsection{Sequential and asynchronous communication} In our current model, all agents communicate synchronously at every time step and update their beliefs accordingly. This simplification is a reasonable approximation of belief update for an initial study of group polarization, but in real social networks, communication may be both asynchronous and  sequential, in the sense that messages are sent one by one between a pair of agents, rather than every agent sending every connected agent a message at every time step as in the current model.   The current model is deterministic, but introducing sequential messages will add nondeterminism to the model, bringing new complications. We plan to develop a logic in the style of dynamic epistemic logic~\cite{del} in order to reason about sequential messages.   Besides being sequential, communication in real social networks is also asynchronous, making it particularly difficult to represent agents' knowledge and beliefs about one another, as discussed in~\cite{async1,async2}.  Eventually, we plan to include asynchronous communication and its effects on beliefs in our model.}

\review{
\subsection{Integrating our approach with data}
The current model is completely theoretical, and in some of the ways mentioned above it is an oversimplification.  Eventually, when our model is more mature, we plan to use data gathered from real social networks both to verify our conclusions and to allow  us to choose the correct parameters in our models (e.g.\ a realistic threshold for when the backfire effect occurs).  Fortunately, there is an increasing amount of empirical work being done on issues such as polarization in social networks~\cite{emp1,emp2,Guerra}. This will make it easier for us to compare our model to the real state of the world and improve it.
}
\review{
\subsection{Applications to issues other than polarization}
Our model of changing beliefs in social networks certainly has other applications besides group polarization. One important issue is the development of false beliefs and the spread of misinformation. In order to explore this problem we again need the notion of outside truth. One of our goals is to learn the conditions under which agents' beliefs tend to move close to truth, and under what conditions false beliefs tend to develop and spread. In particular, we hope to understand whether there are certain classes of social networks which are resistant to the spread of false belief due to their topology and influence conditions.
}

\subsection{Process Calculus} We plan to develop a process calculus which incorporates structures from social networks, such as communication, influence, priority, publicly posted messages, and individual opinions and beliefs. Since process calculi are an efficient, powerful, well studied formalism for reasoning about concurrent multi-agent systems, using them to represent social networks would simplify our models and give us access to already developed tools. In~\cite{concur12,haar:hal-01256984,guzman:hal-02172415,guzman:hal-01257113,ArandaVV09} we developed  process calculi and information systems to reason about the evolution of agents' beliefs and knowledge. Nevertheless, these works do not address the notion of polarization or consensus among the agents of a system.

\subsubsection*{Acknowledgments} 
We would like to thank Benjamin Golub for insightful remarks and referring us to important related work.

\bibliographystyle{alphaurl}
\bibliography{polar}

\newcommand{\etalchar}[1]{$^{#1}$}
\begin{thebibliography}{AvDGvdH14b}

\bibitem[AAK{\etalchar{+}}21]{website:github-repo}
M{\'{a}}rio~S. Alvim, Bernardo Amorim, Sophia Knight, Santiago Quintero, and
  Frank Valencia.
\newblock {Polarization Simulator}.
\newblock \url{https://github.com/Sirquini/Polarization}, 2021.

\bibitem[AGvdH15]{gossip1}
Krzysztof~R. Apt, Davide Grossi, and Wiebe van~der Hoek.
\newblock {Epistemic Protocols for Distributed Gossiping}.
\newblock In {\em Proceedings of the 15th Conference on Theoretical Aspects of
  Rationality and Knowledge (TARK 2015)}, volume 215 of {\em {EPTCS}}, pages
  51--66, 2015.
\newblock \href {https://doi.org/10.4204/EPTCS.215.5}
  {\path{doi:10.4204/EPTCS.215.5}}.

\bibitem[AGvdH18]{gossip2}
Krzysztof~R. Apt, Davide Grossi, and Wiebe van~der Hoek.
\newblock {When Are Two Gossips the Same? Types of Communication in Epistemic
  Gossip Protocols}.
\newblock {\em CoRR}, abs/1807.05283, 2018.
\newblock \href {http://arxiv.org/abs/1807.05283} {\path{arXiv:1807.05283}}.

\bibitem[AKV19]{Alvim:19:FC}
M{\'{a}}rio~S. Alvim, Sophia Knight, and Frank Valencia.
\newblock Toward a formal model for group polarization in social networks.
\newblock In {\em The Art of Modelling Computational Systems: {A} Journey from
  Logic and Concurrency to Security and Privacy - Essays Dedicated to Catuscia
  Palamidessi on the Occasion of Her 60th Birthday}, volume 11760 of {\em
  LNCS}, pages 419--441. Springer, 2019.
\newblock \href {https://doi.org/10.1007/978-3-030-31175-9\_24}
  {\path{doi:10.1007/978-3-030-31175-9\_24}}.

\bibitem[AvDGvdH14a]{gossip3}
Maduka Attamah, Hans van Ditmarsch, Davide Grossi, and Wiebe van~der Hoek.
\newblock {A Framework for Epistemic Gossip Protocols}.
\newblock In {\em Proceedings of the 12th European Conference on Multi-Agent
  Systems (EUMAS 2014)}, volume 8953 of {\em LNCS}, pages 193--209. Springer,
  2014.
\newblock \href {https://doi.org/10.1007/978-3-319-17130-2\_13}
  {\path{doi:10.1007/978-3-319-17130-2\_13}}.

\bibitem[AvDGvdH14b]{gossip4}
Maduka Attamah, Hans van Ditmarsch, Davide Grossi, and Wiebe van~der Hoek.
\newblock {Knowledge and Gossip}.
\newblock In {\em {Proceedings of the 21st European Conference on Artificial
  Intelligence (ECAI 2014)}}, volume 263 of {\em Frontiers in Artificial
  Intelligence and Applications}, pages 21--26. {IOS} Press, 2014.
\newblock \href {https://doi.org/10.3233/978-1-61499-419-0-21}
  {\path{doi:10.3233/978-1-61499-419-0-21}}.

\bibitem[AVV09]{ArandaVV09}
Jes{\'{u}}s Aranda, Frank~D. Valencia, and Cristian Versari.
\newblock {On the Expressive Power of Restriction and Priorities in {CCS} with
  Replication}.
\newblock In {\em {Proceedings of the 12th International Conference on
  Foundations of Software Science and Computational Structures (FOSSACS
  2009)}}, volume 5504 of {\em LNCS}, pages 242--256. Springer, 2009.
\newblock \href {https://doi.org/10.1007/978-3-642-00596-1\_18}
  {\path{doi:10.1007/978-3-642-00596-1\_18}}.

\bibitem[AWA10]{Aronson10}
Elliot Aronson, Timothy Wilson, and Robin Akert.
\newblock {\em Social Psychology}.
\newblock Upper Saddle River, NJ : Prentice Hall, 7 edition, 2010.

\bibitem[BAB{\etalchar{+}}18]{emp1}
Christopher~A. Bail, Lisa~P. Argyle, Taylor~W. Brown, John~P. Bumpus, Haohan
  Chen, M.~B.~Fallin Hunzaker, Jaemin Lee, Marcus Mann, Friedolin Merhout, and
  Alexander Volfovsky.
\newblock {Exposure to opposing views on social media can increase political
  polarization}.
\newblock {\em Proceedings of the National Academy of Sciences},
  115(37):9216--9221, 2018.
\newblock \href {https://doi.org/10.1073/pnas.1804840115}
  {\path{doi:10.1073/pnas.1804840115}}.

\bibitem[BCRS19]{bcrs}
Alexandru Baltag, Zo{\'{e}} Christoff, Rasmus~K. Rendsvig, and Sonja Smets.
\newblock {Dynamic Epistemic Logics of Diffusion and Prediction in Social
  Networks}.
\newblock {\em Stud Logica}, 107(3):489--531, 2019.
\newblock \href {https://doi.org/10.1007/s11225-018-9804-x}
  {\path{doi:10.1007/s11225-018-9804-x}}.

\bibitem[BM68]{nolan}
Maurice~C. Bryson and William~R. McDill.
\newblock {The Political Spectrum: A Bi-Dimensional Approach}.
\newblock {\em Rampart Journal of Individualist Thought}, pages 19--26, 1968.

\bibitem[Boz13]{Bozdag13}
Engin Bozdag.
\newblock Bias in algorithmic filtering and personalization.
\newblock {\em Ethics and Information Technology}, 15(3):209--227, 2013.
\newblock \href {https://doi.org/10.1007/s10676-013-9321-6}
  {\path{doi:10.1007/s10676-013-9321-6}}.

\bibitem[CGMJCK13]{Guerra}
P.H. Calais~Guerra, Wagner Meira~Jr, Claire Cardie, and R~Kleinberg.
\newblock A measure of polarization on social media networks based on community
  boundaries.
\newblock {\em Proceedings of the 7th International Conference on Weblogs and
  Social Media (ICWSM 2013)}, pages 215--224, 01 2013.
\newblock \href {https://doi.org/10.1609/icwsm.v7i1.14421}
  {\path{doi:10.1609/icwsm.v7i1.14421}}.

\bibitem[Chr16]{zoe}
Zo{\'e} Christoff.
\newblock {\em Dynamic logics of networks: information flow and the spread of
  opinion}.
\newblock PhD thesis, PhD Thesis, Institute for Logic, Language and
  Computation, University of Amsterdam, 2016.

\bibitem[CVCL20]{robust}
Simone Cerreia-Vioglio, Roberto Corrao, and Giacomo Lanzani.
\newblock {Robust Opinion Aggregation and its Dynamics}.
\newblock Working Papers 662, IGIER (Innocenzo Gasparini Institute for Economic
  Research), Bocconi University, 2020.

\bibitem[DeG74]{degroot}
Morris~H. DeGroot.
\newblock Reaching a consensus.
\newblock {\em Journal of the American Statistical Association},
  69(345):118--121, 1974.
\newblock \href {https://doi.org/10.2307/2285509} {\path{doi:10.2307/2285509}}.

\bibitem[Die17]{Diestel:17}
Reinhard Diestel.
\newblock {\em Graph Theory}.
\newblock Graduate Texts in Mathematics. Springer-Verlag, 5th edition, 2017.
\newblock \href {https://doi.org/10.1007/978-3-662-53622-3}
  {\path{doi:10.1007/978-3-662-53622-3}}.

\bibitem[DVZ03]{demarzo}
Peter~M DeMarzo, Dimitri Vayanos, and Jeffrey Zwiebel.
\newblock {Persuasion Bias, Social Influence, and Unidimensional Opinions}.
\newblock {\em {The Quarterly Journal of Economics}}, 118(3):909--968, 2003.
\newblock \href {https://doi.org/10.1162/00335530360698469}
  {\path{doi:10.1162/00335530360698469}}.

\bibitem[Eld19]{alexis}
Alexis Elder.
\newblock {The interpersonal is political: unfriending to promote civic
  discourse on social media}.
\newblock {\em Ethics and Information Technology}, pages 1--10, 2019.
\newblock \href {https://doi.org/10.1007/s10676-019-09511-4}
  {\path{doi:10.1007/s10676-019-09511-4}}.

\bibitem[ER94]{Esteban:94:Econometrica}
Joan-Mar\'ia Esteban and Debraj Ray.
\newblock {On the Measurement of Polarization}.
\newblock {\em Econometrica}, 62(4):819--851, 1994.
\newblock \href {https://doi.org/10.2307/2951734} {\path{doi:10.2307/2951734}}.

\bibitem[FGR16]{filter}
Seth Flaxman, Sharad Goel, and Justin~M Rao.
\newblock Filter bubbles, echo chambers, and online news consumption.
\newblock {\em Public opinion quarterly}, 80(S1):298--320, 2016.

\bibitem[GG16]{gargiulo}
Floriana Gargiulo and Yerali Gandica.
\newblock The role of homophily in the emergence of opinion controversies.
\newblock {\em CoRR}, abs/1612.05483, 2016.
\newblock \href {http://arxiv.org/abs/1612.05483} {\path{arXiv:1612.05483}}.

\bibitem[GHP{\etalchar{+}}16]{guzman:hal-01257113}
Michell Guzm{\'a}n, Stefan Haar, Salim Perchy, Camilo Rueda, and Frank
  Valencia.
\newblock {Belief, Knowledge, Lies and Other Utterances in an Algebra for Space
  and Extrusion}.
\newblock {\em {Journal of Logical and Algebraic Methods in Programming}},
  September 2016.
\newblock \href {https://doi.org/10.1016/j.jlamp.2016.09.001}
  {\path{doi:10.1016/j.jlamp.2016.09.001}}.

\bibitem[GJ10]{naive}
Benjamin Golub and Matthew~O. Jackson.
\newblock {Na\"ive Learning in Social Networks and the Wisdom of Crowds}.
\newblock {\em American Economic Journal: Microeconomics}, 2(1):112--49, 2010.
\newblock \href {https://doi.org/10.1257/mic.2.1.112}
  {\path{doi:10.1257/mic.2.1.112}}.

\bibitem[GKQ{\etalchar{+}}19]{guzman:hal-02172415}
Michell Guzm{\'a}n, Sophia Knight, Santiago Quintero, Sergio Ram{\'i}rez,
  Camilo Rueda, and Frank~D. Valencia.
\newblock {Reasoning about Distributed Knowledge of Groups with Infinitely Many
  Agents}.
\newblock In {\em Proceedings of the 30th International Conference on
  Concurrency Theory (CONCUR 2019)}, volume~29, pages 1--29, August 2019.
\newblock \href {https://doi.org/10.4230/LIPIcs.CONCUR.2019.29}
  {\path{doi:10.4230/LIPIcs.CONCUR.2019.29}}.

\bibitem[GS16]{survey}
Benjamin Golub and Evan Sadler.
\newblock {Learning in Social Networks}.
\newblock In {\em {The Oxford Handbook of the Economics of Networks}}. Oxford
  University Press, 04 2016.
\newblock \href {https://doi.org/10.1093/oxfordhb/9780199948277.013.12}
  {\path{doi:10.1093/oxfordhb/9780199948277.013.12}}.

\bibitem[HK02]{hk}
Rainer Hegselmann and Ulrich Krause.
\newblock {Opinion Dynamics and Bounded Confidence, Models, Analysis and
  Simulation}.
\newblock {\em Journal of Artificial Societies and Social Simulation}, 5(3):2,
  2002.
\newblock URL: \url{https://www.jasss.org/5/3/2.html}.

\bibitem[HPRV15]{haar:hal-01256984}
Stefan Haar, Salim Perchy, Camilo Rueda, and Frank Valencia.
\newblock {An Algebraic View of Space/Belief and Extrusion/Utterance for
  Concurrency/Epistemic Logic}.
\newblock In {\em Proceedings of the 17th International Symposium on Principles
  and Practice of Declarative Programming (PPDP 2015)}, pages 161--172. {ACM
  SIGPLAN}, July 2015.
\newblock \href {https://doi.org/10.1145/2790449.2790520}
  {\path{doi:10.1145/2790449.2790520}}.

\bibitem[HRW08]{huberman}
Bernardo~A Huberman, Daniel~M Romero, and Fang Wu.
\newblock Social networks that matter: Twitter under the microscope.
\newblock {\em CoRR}, abs/0812.1045, 2008.
\newblock \href {http://arxiv.org/abs/0812.1045} {\path{arXiv:0812.1045}}.

\bibitem[Hun17]{hunter}
Aaron Hunter.
\newblock {Reasoning About Trust and Belief Change on a Social Network: A
  Formal Approach}.
\newblock In {\em Proceedings of the 13th International Conference on
  Information Security Practice and Experience (ISPEC 2017)}, pages 783--801.
  Springer, 2017.
\newblock \href {https://doi.org/10.1007/978-3-319-72359-4_49}
  {\path{doi:10.1007/978-3-319-72359-4_49}}.

\bibitem[Kir17]{Kirby17}
E.J. Kirby.
\newblock The city getting rich from fake news.
\newblock BBC News Documentary, 05 2017.
\newblock URL: \url{https://www.bbc.com/news/magazine-38168281}.

\bibitem[KLCKK14]{emp2}
Jae Kook~Lee, Jihyang Choi, Cheonsoo Kim, and Yonghwan Kim.
\newblock {Social Media, Network Heterogeneity, and Opinion Polarization}.
\newblock {\em Journal of Communication}, 64, 08 2014.
\newblock \href {https://doi.org/10.1111/jcom.12077}
  {\path{doi:10.1111/jcom.12077}}.

\bibitem[KMS15]{async1}
Sophia Knight, Bastien Maubert, and Fran{\c{c}}ois Schwarzentruber.
\newblock {Asynchronous Announcements in a Public Channel}.
\newblock In {\em Proceedings of the 12th International Colloquium on
  Theoretical Aspects of Computing (ICTAC 2015)}, pages 272--289, 2015.
\newblock \href {https://doi.org/10.1007/978-3-319-25150-9_17}
  {\path{doi:10.1007/978-3-319-25150-9_17}}.

\bibitem[KMS19]{async2}
Sophia Knight, Bastien Maubert, and Fran{\c{c}}ois Schwarzentruber.
\newblock Reasoning about knowledge and messages in asynchronous multi-agent
  systems.
\newblock {\em Mathematical Structures in Computer Science}, 29(1):127--168,
  2019.
\newblock \href {https://doi.org/10.1017/S0960129517000214}
  {\path{doi:10.1017/S0960129517000214}}.

\bibitem[KPPV12]{concur12}
Sophia Knight, Catuscia Palamidessi, Prakash Panangaden, and Frank~D. Valencia.
\newblock {Spatial and Epistemic Modalities in Constraint-Based Process
  Calculi}.
\newblock In {\em Proceedings of the 23rd International Conference on
  Concurrency Theory (CONCUR 2012)}, pages 317--332, 2012.
\newblock \href {https://doi.org/10.1007/978-3-642-32940-1_23}
  {\path{doi:10.1007/978-3-642-32940-1_23}}.

\bibitem[LSG14]{liu}
Fenrong Liu, Jeremy Seligman, and Patrick Girard.
\newblock Logical dynamics of belief change in the community.
\newblock {\em Synthese}, 191(11):2403--2431, Jul 2014.
\newblock \href {https://doi.org/10.1007/s11229-014-0432-3}
  {\path{doi:10.1007/s11229-014-0432-3}}.

\bibitem[LSSZ13]{li}
Lin Li, Anna Scaglione, Ananthram Swami, and Qing Zhao.
\newblock Consensus, polarization and clustering of opinions in social
  networks.
\newblock {\em IEEE Journal on Selected Areas in Communications},
  31(6):1072--1083, 2013.
\newblock \href {https://doi.org/10.1109/JSAC.2013.130609}
  {\path{doi:10.1109/JSAC.2013.130609}}.

\bibitem[Lyn96]{Lynch96}
Nancy~A. Lynch.
\newblock {\em Distributed Algorithms}.
\newblock The Morgan Kaufmann Series in Data Management Systems. Morgan
  Kaufmann Publishers, 1996.

\bibitem[MBA20]{mao}
Y.~{Mao}, S.~{Bolouki}, and E.~{Akyol}.
\newblock {Spread of Information With Confirmation Bias in Cyber-Social
  Networks}.
\newblock {\em IEEE Transactions on Network Science and Engineering},
  7(2):688--700, 2020.
\newblock \href {https://doi.org/10.1109/TNSE.2018.2878377}
  {\path{doi:10.1109/TNSE.2018.2878377}}.

\bibitem[MF15]{mf}
Manuel Mueller-Frank.
\newblock {Reaching Consensus in Social Networks}.
\newblock IESE Research Papers D/1116, IESE Business School, February 2015.
\newblock \href {https://doi.org/10.2139/ssrn.2693704}
  {\path{doi:10.2139/ssrn.2693704}}.

\bibitem[ML76]{M76}
D.~G Myers and H.~Lamm.
\newblock {The Group Polarization Phenomenon}.
\newblock {\em Psychological Bulletin}, 1976.
\newblock \href {https://doi.org/10.1037/0033-2909.83.4.602}
  {\path{doi:10.1037/0033-2909.83.4.602}}.

\bibitem[{Mor}05]{moreau}
L.~{Moreau}.
\newblock Stability of multiagent systems with time-dependent communication
  links.
\newblock {\em IEEE Transactions on Automatic Control}, 50(2):169--182, 2005.
\newblock \href {https://doi.org/10.1109/TAC.2004.841888}
  {\path{doi:10.1109/TAC.2004.841888}}.

\bibitem[NPV02]{ntcc}
Mogens Nielsen, Catuscia Palamidessi, and Frank~D. Valencia.
\newblock {Temporal Concurrent Constraint Programming: Denotation, Logic and
  Applications}.
\newblock {\em Nordic Journal of Computing}, 9(1):145--188, 2002.
\newblock URL: \url{https://dl.acm.org/doi/abs/10.5555/643009.643014}.

\bibitem[Ped]{myp}
Mina~Young Pedersen.
\newblock {Polarization and Echo Chambers: A Logical Analysis of Balance and
  Triadic Closure in Social Networks}.
\newblock URL: \url{https://eprints.illc.uva.nl/id/eprint/1700/}.

\bibitem[Pla07]{del}
Jan Plaza.
\newblock Logics of public communications.
\newblock {\em Synthese}, 158(2):165--179, 2007.
\newblock \href {https://doi.org/10.1007/s11229-007-9168-7}
  {\path{doi:10.1007/s11229-007-9168-7}}.

\bibitem[PMC16]{proskurnikov}
A.~V. {Proskurnikov}, A.~S. {Matveev}, and M.~{Cao}.
\newblock {Opinion Dynamics in Social Networks With Hostile Camps: Consensus
  vs. Polarization}.
\newblock {\em IEEE Transactions on Automatic Control}, 61(6):1524--1536, June
  2016.
\newblock \href {https://doi.org/10.1109/TAC.2015.2471655}
  {\path{doi:10.1109/TAC.2015.2471655}}.

\bibitem[PS{\AA}19]{myp2}
Mina~Young Pedersen, Sonja Smets, and Thomas {\AA}gotnes.
\newblock {Analyzing Echo Chambers: A Logic of Strong and Weak Ties}.
\newblock In {\em Proceedings of the 7th International Workshop on Logic,
  Rationality and Interaction (LORI 2019)}, pages 183--198. Springer, 2019.
\newblock \href {https://doi.org/10.1007/978-3-662-60292-8_14}
  {\path{doi:10.1007/978-3-662-60292-8_14}}.

\bibitem[PS{\AA}20]{myp3}
Mina~Young Pedersen, Sonja Smets, and Thomas {\AA}gotnes.
\newblock {Further Steps Towards a Logic of Polarization in Social Networks}.
\newblock In {\em Logic and Argumentation}, pages 324--345. Springer
  International Publishing, 2020.
\newblock \href {https://doi.org/10.1007/978-3-030-44638-3_20}
  {\path{doi:10.1007/978-3-030-44638-3_20}}.

\bibitem[Ram19]{Ramos:19:Book}
Ver\'{o}nica~Ju\'{a}rez Ramos.
\newblock {\em {Analyzing the Role of Cognitive Biases in the Decision-Making
  Process}}.
\newblock IGI Global, 2019.
\newblock \href {https://doi.org/10.4018/978-1-5225-2978-1}
  {\path{doi:10.4018/978-1-5225-2978-1}}.

\bibitem[SJG94]{tcc}
Vijay~A. Saraswat, Radha Jagadeesan, and Vineet Gupta.
\newblock Foundations of timed concurrent constraint programming.
\newblock In {\em {Proceedings the 9th Symposium on Logic in Computer Science
  (LICS 1994)}}, pages 71--80. {IEEE}, 1994.
\newblock \href {https://doi.org/10.1109/LICS.1994.316085}
  {\path{doi:10.1109/LICS.1994.316085}}.

\bibitem[SLG11]{fblogic}
Jeremy Seligman, Fenrong Liu, and Patrick Girard.
\newblock Logic in the community.
\newblock In {\em Indian Conference on Logic and Its Applications (ICLA 2011)},
  pages 178--188. Springer, 2011.
\newblock \href {https://doi.org/10.1007/978-3-642-18026-2_15}
  {\path{doi:10.1007/978-3-642-18026-2_15}}.

\bibitem[SLG13]{facebook}
Jeremy Seligman, Fenrong Liu, and Patrick Girard.
\newblock Facebook and the epistemic logic of friendship.
\newblock {\em CoRR}, abs/1310.6440, 2013.
\newblock \href {http://arxiv.org/abs/1310.6440} {\path{arXiv:1310.6440}}.

\bibitem[Soh14]{Sohrab:14}
Houshang~H. Sohrab.
\newblock {\em Basic Real Analysis}.
\newblock Birkhauser Basel, 2nd edition, 2014.
\newblock \href {https://doi.org/10.1007/0-8176-4441-5}
  {\path{doi:10.1007/0-8176-4441-5}}.

\bibitem[SPGK18]{sirbu}
Alina S{\^\i}rbu, Dino Pedreschi, Fosca Giannotti, and J{\'a}nos Kert{\'e}sz.
\newblock Algorithmic bias amplifies opinion polarization: A bounded confidence
  model.
\newblock {\em CoRR}, abs/1803.02111, 2018.
\newblock \href {http://arxiv.org/abs/1803.02111} {\path{arXiv:1803.02111}}.

\bibitem[SSVL20]{sikder}
Orowa Sikder, Robert Smith, Pierpaolo Vivo, and Giacomo Livan.
\newblock A minimalistic model of bias, polarization and misinformation in
  social networks.
\newblock {\em Scientific Reports}, 10, 03 2020.
\newblock \href {https://doi.org/10.1038/s41598-020-62085-w}
  {\path{doi:10.1038/s41598-020-62085-w}}.

\bibitem[Val01]{Valencia01}
Frank~D. Valencia.
\newblock {Temporal Concurrent Constraint Programming}.
\newblock In {\em Proceedings of the 7th International Conference on Principles
  and Practice of Constraint Programming (CP 2001)}, volume 2239 of {\em LNCS},
  page 786. Springer, 2001.
\newblock \href {https://doi.org/10.1007/3-540-45578-7\_84}
  {\path{doi:10.1007/3-540-45578-7\_84}}.

\end{thebibliography}

\newpage
\appendix
\section{Axioms for Esteban-Ray polarization measure}%
\label{sec:polar-axioms}

The Esteban-Ray polarization measure used in this paper was developed as the only function (up to constants $\alpha$ and $K$) satisfying all of the following conditions and axioms~\cite{Esteban:94:Econometrica}:
\begin{description}
   \item[Condition H]
    The ranking induced by the polarization measure over two
    distributions is invariant \review{with respect to} the size of the population:~\footnote{This is why we can assume \review{without loss of generality} that the distribution is a probability distribution.}
    \[
     \PfunER{\pi, y} \geq \PfunER{\pi', y'} \quad \rightarrow \quad \forall \lambda > 0, \,\, \PfunER{\lambda\pi, y} \geq \PfunER{\lambda\pi', y'}~.
    \]

    \item[Axiom 1]
    Consider three levels of belief $p, q, r\in[0,1]$
    such that the same proportion of the population holds beliefs $q$ and $r$, and
    a significantly higher proportion of the population holds belief $p$.
    If the groups of agents that hold beliefs $q$ and $r$ reach a consensus and
    agree on an \qm{average} belief \review{$\nicefrac{(q+r)}{2}$}, then the social network becomes
    more polarized.

    \item[Axiom 2]
    Consider three levels of belief $p, q, r \in [0,1]$,
    such that $q$ is at least as close to $r$ as it is to $p$, and
    $p>r$.
    If only small variations on $q$ are permitted, the direction that brings it closer
    to the nearer and smaller opinion ($r$) should increase polarization.

    \item[Axiom 3]
    Consider three levels of belief $p, q, r \in [0,1]$, \review{such that} $p<q<r$ and there
    is a non-zero proportion of the population holding belief $q$.
    If the proportion of the population that holds belief $q$ is equally split into
    holding beliefs $q$ and $r$, then polarization increases.
\end{description}



\section{Proofs}%
\label{sec:proofs}


\respolatlimit*

\proof
Let  be any real $\epsilon>0$. It suffices to find $N \in \reals$  such that for every $t>N$, \\$Pfun{\Blft{t}} < \epsilon.$
Let $I_m$ be the bin of $D_k$ such that $v \in I_m.$ Let $l$ and $r$ be the left and right  endpoints of $I_m$, respectively.

Take
\[\epsilon' =
\begin{cases}
r & \textrm{ if }v=0,\\
l & \textrm{ if }v=1,\\
\min\{ v - l, r-v\} & \textrm{otherwise}.
\end{cases}
\]
Clearly $\epsilon' >0$ because $v$ is not a borderline point. Since \[\lim_{t \to \infty} \Bfun{i}{t} = v,\] there is some $N_i \in \reals$  such that for every $t>N_i$, $|v - \Bfun{i}{t}| < \epsilon'$.  This implies that  $\Bfun{i}{t} \in I_m$ for every $t>N_i$.  Take \[N= \max\{ N_i | i \in \Agents \}.\]  From Proposition~\ref{pol-consensus} $\Pfun{\Blft{t}}=0 < \epsilon$ for every $t>N$,  as wanted. \qed

\reslemmacbmaxdiffmin*

\proof
We want to prove that \[\Bfun{\agent{i}}{t{+}1} \leq \mx{t}.\] Since \[\Bfun{\agent{j}}{t} \leq \mx{t},\] we can use Definition~\ref{def:confirmation-bias}
to derive the inequality \[\Bfun{\agent{i}}{t{+}1}
    \leq E_1 \defsymbol \Bfun{\agent{i}}{t} + \frac{1}{|\Agents_i|}\sum_{\agent{j} \in \Agents_i \setminus \{\agent{i}\}}  \CBfun{\agent{i}}{\agent{j}}{t}\Ifun{\agent{j}}{\agent{i}}(\mx{t} -\Bfun{\agent{i}}{t}).\]
    Furthermore,
    \[E_1 \leq E_2 \defsymbol\Bfun{\agent{i}}{t} + \frac{1}{|\Agents_i|}\sum_{\agent{j} \in \Agents_i \setminus \{\agent{i}\}} (\mx{t} -\Bfun{\agent{i}}{t})\]
    because
    \[\CBfun{\agent{i}}{\agent{j}}{t}\Ifun{\agent{j}}{\agent{i}} \leq 1\]
    and
    \[\mx{t} - \Bfun{\agent{i}}{t} \geq 0.\]
    We thus obtain
    \begin{eqnarray*}
    \Bfun{\agent{i}}{t{+}1}
    &\leq&
    E_2\\
    &=&
    \Bfun{\agent{i}}{t} + \frac{|\Agents_i|-1}{|\Agents_i|}(\mx{t} -\Bfun{\agent{i}}{t}) \\
    &=&
    \frac{\Bfun{\agent{i}}{t} + (|\Agents_i|-1)\cdot\mx{t}}{|\Agents_i|}\\ &\leq &
    \mx{t}
    \end{eqnarray*}
    as wanted.

    The proof that $\mn{t} \leq \Bfun{\agent{i}}{t{+}1}$ is similar.
\qed


\begin{restatable}[]{prop}{respropcbupperboundinequality}%
\label{prop:cb-upper-bound-inequality}
Let $i \in \Agents$, $k \in \Agents_i$, $n,t \in \nat$ with $n\geq 1$, and $v \in [0,1]$.
  \begin{enumerate}
        \item  If $\Bfun{\agent{i}}{t} \leq v$ then
            \[\Bfun{\agent{i}}{t{+}1} \leq v + \frac{1}{|\Agents|}\sum_{\agent{j} \in \Agents_i}\CBfun{i}{j}{t}\Ifun{\agent{j}}{\agent{i}}\left(\Bfun{\agent{j}}{t}-v\right).\]
        \item It is always the case that \[\Bfun{\agent{i}}{t+n} \leq \mx{t} + \frac{1}{|\Agents|} \, \CBfun{i}{k}{t+n-1}\Ifun{\agent{k}}{\agent{i}}(\Bfun{\agent{k}}{t+n-1} - \mx{t}).\]
     \end{enumerate}
\end{restatable}

\proof Let $i,k \in \Agents$, $n,t \in \nat$ with $n\geq 1$, and $v \in [0,1]$.
\begin{enumerate}
\item From Definition~\ref{def:confirmation-bias}:
\begin{eqnarray*}
\Bfun{\agent{i}}{t{+}1}
&=&
\Bfun{\agent{i}}{t} + \frac{1}{|\Agents_i|}\sum_{\agent{j} \in \Agents_i}\CBfun{i}{j}{t}\Ifun{\agent{j}}{\agent{i}}(\Bfun{\agent{j}}{t} - \Bfun{\agent{i}}{t})\\
&\leq&
v +  \frac{1}{|\Agents_i|}\sum_{\agent{j} \in \Agents_i \setminus \{\agent{i}\}}\CBfun{i}{j}{t}\Ifun{\agent{j}}{\agent{i}}(\Bfun{\agent{j}}{t} - v) \\
&\leq&
v + \frac{1}{|\Agents|}\sum_{\agent{j} \in \Agents_i\setminus \{\agent{i}\}}\CBfun{i}{j}{t}\Ifun{\agent{j}}{\agent{i}}(\Bfun{\agent{j}}{t} - v) \\
&=&
v + \frac{1}{|\Agents|}\sum_{\agent{j} \in \Agents_i}\CBfun{i}{j}{t}\Ifun{\agent{j}}{\agent{i}}(\Bfun{\agent{j}}{t} - v)
\end{eqnarray*}
since $|\Agents_i| \leq |\Agents|$.
\item From Proposition~\ref{prop:cb-upper-bound-inequality}(1): 
\begin{eqnarray*}
\Bfun{\agent{i}}{t+n}
&\leq&
\mx{t} + \frac{1}{|\Agents|}\sum_{\agent{j} \in \Agents_i}\CBfun{i}{j}{t+n-1}\Ifun{\agent{j}}{\agent{i}}\left(\Bfun{\agent{j}}{t+n-1}-\mx{t}\right) \\
&\leq &
\mx{t} + \frac{1}{|\Agents|}\CBfun{i}{k}{t+n-1}\Ifun{\agent{k}}{\agent{i}}\left(\Bfun{\agent{k}}{t+n-1}-\mx{t}\right)
\end{eqnarray*}
using Corollary~\ref{cor:biasfactor} and the fact that \[\Bfun{\agent{j}}{t+n-1}-\mx{t} \leq 0. \qedhere \]
\end{enumerate}

\reslemmacbpathbound*

\proof
\begin{enumerate}
    \item Let $p$ be the path
     \[i_0\to^{C_1}i_1\to^{C_2}\ldots\to^{C_n}i_n.\]
      We proceed by induction on $n$. For $n=1$, since \[\Bfun{\agent{i_0}}{t} - \mx{t} \leq 0,\]  we obtain the result immediately from Proposition~\ref{prop:cb-upper-bound-inequality}(2) and Proposition~\ref{fcb-min:prop}. 
      Assume that  \[\Bfun{\agent{i_{n-1}}}{t+|p'|} \leq \mx{t} + \frac{C'\CBfunM^{|p'|}}{|\Agents|^{|p'|}}(
    \Bfun{\agent{i_0}}{t} - \mx{t})\]  where
    \[p' = i_0\to^{C_1}i_1\to^{C_2}\ldots\to^{C_{n-1}}i_{n-1}\]
    and
    \[C' =C_1\times \cdots \times C_{n-1}\]
    with $n> 1$. Notice that $|p'|=|p|-1$.
    Using Proposition~\ref{prop:cb-upper-bound-inequality}(2), Proposition~\ref{fcb-min:prop}, and the fact that 
    \[\Bfun{\agent{i_{n-1}}}{t+|p'|} - \mx{t} \leq 0\]
    we obtain
    \[\Bfun{\agent{i_n}}{t+|p|} \leq \mx{t} + \frac{C_n\CBfunM}{|\Agents|}(\Bfun{\agent{i_{n-1}     }}{t+|p'|} - \mx{t}).\]
    Using our assumption
    we obtain
    \begin{eqnarray*}
    \Bfun{\agent{i_n}}{t+|p|}
    & \leq&
    \mx{t} + \frac{C_n\CBfunM}{|\Agents|}( \mx{t} + \frac{C'\CBfunM^{|p'|}}{|\Agents|^{|p'|}}(
    \Bfun{\agent{i_0}}{t} - \mx{t})  - \mx{t})\\
    &=&
    \mx{t} + \frac{C\CBfunM^{|p|}}{|\Agents|^{|p|}}(
    \Bfun{\agent{i_0}}{t} - \mx{t})
    \end{eqnarray*}
    as wanted.
    \item Suppose that $p$ is the path
    \[\linfl{\agent{\mstar}^t}{C}{p}{\agent{i}}.\]
    From Lemma~\ref{lemma:cb-path-bound}(1) we obtain 
    \begin{eqnarray*}
    \Bfun{\agent{i}}{t+|p|}
    &\leq&
    \mx{t} + \frac{C\CBfunM^{|p|}}{|\Agents|^{|p|}}(\Bfun{\agent{\mstar}^t}{t} - \mx{t})\\
    &=&
    \mx{t} + \frac{C\CBfunM^{|p|}}{|\Agents|^{|p|}}(\mn{t} - \mx{t}).
    \end{eqnarray*}
    Since
    \[\frac{C\CBfunM^{|p|}}{|\Agents|^{|p|}}(\mn{t} - \mx{t}) \leq 0,\]
    we can substitute $\IfunM^{|p|}$ for $C$. Thus,
    \[\Bfun{\agent{i}}{t+|p|} \leq \mx{t} + \Big(\frac{\IfunM\CBfunM}{|\Agents|}\Big)^{|p|} \allowbreak (\mn{t}-\mx{t}).\]
    From Theorem~\ref{th:cb-max-limits-exist}, the maximum value of $\mn{t}$ is $L$ and the minimum value of $\mx{t}$ is $U$, thus
    \begin{align*}
    \Bfun{\agent{i}}{t+|p|}
    &\leq
    \mx{t} + \Big(\frac{\IfunM\CBfunM}{|\Agents|}\Big)^{|p|}(L-U)\\
    &=
    \mx{t} - \delta. \qedhere
    \end{align*}
\end{enumerate}


\reslemcbepsilonbound*

\proof
\begin{enumerate}
    \item Using Proposition~\ref{prop:cb-upper-bound-inequality}(1) with the assumption that \[\Bfun{\agent{i}}{t{+}n} \leq \mx{t} - \gamma\] 
    for $\gamma \geq 0$ and the fact that $\Ifun{\agent{j}}{\agent{i}} \in [0,1]$  we obtain the inequality
    \[\Bfun{\agent{i}}{t+n+1} \leq   \mx{t} - \gamma + \frac{1}{|\Agents|}\sum_{\agent{j} \in \Agents_i}\CBfun{i}{j}{t+n}\Ifun{\agent{j}}{\agent{i}}\left(\Bfun{\agent{j}}{t{+}n} - (\mx{t} - \gamma)\right).\]
    From Corollary~\ref{cor:cb-mbefore-mafter},  $\mx{t}\geq \mx{t+n} \geq \Bfun{\agent{j}}{t+n} $ for every $\agent{j}\in \Agents$, hence
    \[\Bfun{\agent{i}}{t+n+1} \leq \mx{t} - \gamma + \frac{1}{|\Agents|}\sum_{\agent{j} \in \Agents_i}\CBfun{\agent{i}}{\agent{j}}{t+n}\Ifun{\agent{j}}{\agent{i}}\big(\mx{t} - (\mx{t} - \gamma)\big).\]
    Since
    \[\CBfun{i}{j}{t+n}\Ifun{\agent{j}}{\agent{i}} \in [0,1],\]  we derive
    \begin{eqnarray*}
    \Bfun{\agent{i}}{t+n+1}
    &\leq&
    \mx{t} - \gamma + \frac{1}{|\Agents|}\sum_{\agent{j} \in \Agents_i}\gamma \\
    &\leq&
    \mx{t} - \frac{\gamma}{|\Agents|}.
    \end{eqnarray*}

    \item Let $p$ be the path $\linfl{\agent{\mstar}^t}{}{p}{\agent{i}}$ where $\mstar^t \in \Agents$ is minimal agent at time $t$ and let
    \[\delta =  \left(\frac{\IfunM\CBfunM}{|\Agents|}\right)^{|p|}(U-L).\]
    If $|p| = |\Agents|-1$ then the result follows from Lemma~\ref{lemma:cb-path-bound}(2). 
    Otherwise $|p| < |\Agents|-1$ by Definition~\ref{def:influence-path}.  We first show by induction on $m$ that
    \[\Bfun{\agent{i}}{t+|p|+m} \leq \mx{t} - \frac{\delta}{|\Agents|^m}\]
    for every $m\geq 0$. If $m=0$, then
    \[\Bfun{\agent{i}}{t+|p|} \leq \mx{t} - \delta,\]
    by Lemma~\ref{lemma:cb-path-bound}(2). 
    If $m>0$ and
    \[\Bfun{\agent{i}}{t+|p|+(m-1)} \leq \mx{t} - \frac{\delta}{|\Agents|^{m-1}}\]
    then
    \[\Bfun{\agent{i}}{t+|p|+m} \leq \mx{t} - \frac{\delta}{|\Agents|^{m}}\]
    by Lemma~\ref{lem:cb-epsilon-bound}(1). 
    Therefore, take $m=|\Agents|-|p|-1$ to obtain \begin{eqnarray*}
    \Bfun{\agent{i}}{t+|\Agents|-1}
    &\leq&
    \mx{t} - \frac{\delta}{|\Agents|^{|\Agents|-|p|-1}}\\
    &=&
    \mx{t} - \frac{(\IfunM\CBfunM)^{|p|}.(U-L)}{|\Agents|^{|\Agents|-1}}\\
    &\leq&
    \mx{t} - \epsilon
    \end{eqnarray*}
    as wanted. \qedhere
\end{enumerate}

\resthul*

\proof
Suppose, by contradiction, that
\[\lim_{t\to\infty} \mx{t}=U \neq L=\lim_{t\to\infty} \mn{t}.\]   Let
\[\epsilon = \left(\frac{\IfunM\CBfunM}{|\Agents|}\right)^{|\Agents|-1}(U-L).\]
From the assumption $U > L$ and Corollary~\ref{cor:biasfactor} we get that $\epsilon > 0$. Take $t=0$ and
\[m=\left(\ceil{\frac{1}{\epsilon}}+1\right).\]
Using
Corollary~\ref{cor:max-dif} we obtain
\[\mx{0}  \geq \mx{{m(|\Agents|-1)}} + m\epsilon.\]
Since $m\epsilon > 1$ and
\[\mx{m(|\Agents|-1)} \geq 0\]
then $\mx{0}> 1$. But this contradicts Definition~\ref{def:extreme:beliefs} which states that $\mx{0} \in [0,1]$.
\qed

%
%


\begin{restatable}[Influencing the Extremes]{prop}{respropcbinfluencingextremes}%
\label{prop:cb-influencing-extremes}
If $\Inter$ is strongly connected and $\Blft{0}$ is not radical, then
\[\mx{|\Agents|{-}1}{<}1.\]
\end{restatable}

\proof Since $\Blft{0}$ is not radical, there must be at least one agent $k$  such that
\[\Bfun{k}{0}\in (0,1).\]  Since $\Inter$ is strongly connected, it suffices to show that for every path
\[\linfl{\agent{k}}{}{p}{\agent{i}},\] we have
\[\Bfun{i}{|p|} < 1.\]
Proceed by induction on size $n$ of the path $p=ki_1\ldots i_{n}$.
For $n = 0$, it is true via the hypothesis. For $n \geq 1$, we have, by IH and Definition~\ref{def:influence-path}, that
\[\Bfun{i_{n-1}}{|p|-1} < 1\]
and
\[\Ifun{i_{n-1}}{n} > 0.\] Thus,
\[\Bfun{i_n}{|p|} = \Bfun{i_n}{|p|-1} + \frac{1}{|\Agents_i|}\sum_{\agent{j} \in \Agents_i} \CBfun{i_n}{j}{|p|-1}\Ifun{j}{i_n} (\Bfun{j}{|p|-1} - \Bfun{i_n}{|p|-1}),\]
and separating $i_{n-1}$ from the sum, we get
\begin{align*}
\Bfun{i_n}{|p|}
&=
\Bfun{i_n}{|p|-1} + \frac{1}{|\Agents_i|}\sum_{\agent{j} \in \Agents_i\setminus\{i_{n-1}\}} \CBfun{i_n}{j}{|p|-1}\Ifun{j}{i_n} (\Bfun{j}{|p|-1} - \Bfun{i_n}{|p|-1}) +\\
&\hspace{1cm}+\frac{1}{|\Agents_i|} \CBfun{i_n}{i_{n-1}}{|p|-1}\Ifun{i_{n-1}}{i_n} (\Bfun{i_{n-1}}{|p|-1} - \Bfun{i_n}{|p|-1}) \\
&\leq 1 + \frac{1}{|\Agents_i|}\CBfun{i_n}{i_{n-1}}{|p|-1}\Ifun{i_{n-1}}{i_n} (\Bfun{i_{n-1}}{|p|-1} - 1)\\
&< 1. \qedhere
\end{align*}

\restheoremcbgeberalsccconvergence*

\proof
\begin{enumerate}
    \item If there exists an agent $\agent{k} \in \Agents$ such that
    \[\Bfun{k}{0} \notin \{0,1\},\]
    then we can use Proposition~\ref{prop:cb-influencing-extremes} to show that by time $|\Agents|-1$, no agent has belief $1$, thus we fall in the general case stated in the beginning of the section (starting at a different time step does not make any difference for these purposes) and thus, all beliefs converge to the same value according to Corollary~\ref{cor:cb-scc-convergence}.
    \item Otherwise it is easy to see that beliefs remain constant as $0$ or $1$ throughout time, since the agents are so biased that the only agents $\agent{j}$ able to influence another agent $\agent{i}$ ($\CBfun{i}{j}{t} \neq 0$) have the same belief as $\agent{i}$. \qedhere
\end{enumerate}

\respropgroupinfluenceconservation*

\proof
Immediate consequence of  Proposition 6.1.1 in~\cite{Diestel:17}.
\qed

\resprocirculationpath*

\proof
For the sake of contradiction, assume  that $\Inter$ is balanced (a circulation) and $\Ifun{i}{j} > 0$ but there is no path from $\agent{j}$ to $\agent{i}$. Define the agents reachable from $j$,
\[R_j = \{k \in
     \Agents | \ \Path{j}{k} \} \cup \{j\}\] and let $\overline{R}_j = \Agents\setminus R_j$. Notice that $\{ R_j ,\overline{R}_j\}$ is a partition of $\Agents$. Since  the codomain of $\Inter$ is $[0,1]$, $\agent{i} \in \overline{R}_j$, $\agent{j} \in R_j$ and $\Ifun{i}{j} > 0$ we obtain
    \[\sum_{k \in R_j}\sum_{l \in \overline{R}_j} \Ifun{l}{k} > 0.\]
    Clearly there is no $k \in R_j, l \in \overline{R}_j$ such that $\Ifun{k}{l} > 0$, therefore
    \[\sum_{k \in R_j}\sum_{l \in \overline{R}_j} \Ifun{k}{l} = 0\]
    which contradicts Proposition~\ref{prop:group-influence-conservation}.
\qed

\respolnonzero*

\proof From Lemma~\ref{prop:circulation-path} it follows that  if the influence graph $\Inter$ is balanced and weakly connected then $\Inter$ is also strongly connected. The result follows from  Lemma~\ref{pol-at-limit} and Theorem~\ref{theorem:cb-geberal-scc-convergence}.
\qed

\rescordegroot*

\proof
    Since the graph is strongly connected it suffices to show that the graph represented by the matrix $P$ in which $P_{i,j}{=}p_{i,j}$ is aperiodic. Since for every individual $i$, $p_{i,i}{>}0$, there is a self-loop, thus no number $K > 1$ divides the length of all cycles in the graph, implying aperiodicity. Thus, the conditions for Theorem 2 of~\cite{degroot} are met, which completes the proof.
\qed

\end{document}